\documentclass[twocolumn]{bmcart}

\usepackage{amsthm, amsmath}
\usepackage[utf8]{inputenc} 

\usepackage{multirow}
\usepackage{pdflscape}
\usepackage{soul}
\usepackage{tikz}
\usetikzlibrary{trees}

\startlocaldefs
\usepackage{graphicx}
\endlocaldefs

\def\checkmark{\tikz\fill[scale=0.3](0,.35) -- (.25,0) -- (1,.7) -- (.25,.15) -- cycle;} 

\begin{document}

\begin{frontmatter}

\begin{fmbox}
\dochead{Research}


\title{Orchestration in the Cloud-to-Things Compute Continuum: Taxonomy, Survey and Future Directions}


\author[
  addressref={aff1, aff2},                   
  corref={aff1},                       
  email={a.ullah@napier.ac.uk}   
]{\fnm{Amjad} \snm{Ullah}}
\author[
 addressref={aff2},
 email={T.Kiss@westminster.ac.uk}
]{\fnm{Tamas} \snm{Kiss}}
\author[
 addressref={aff2, aff3},
 email={j.kovacs@westminster.ac.uk, jozsef.kovacs@sztaki.hu}
]{\fnm{József} \snm{Kovács}}
\author[
 addressref={aff2, aff4},
 email={F.Tusa@westminster.ac.uk}
]{\fnm{Francesco} \snm{Tusa}}
\author[
 addressref={aff2},
 email={J.Deslauriers@westminster.ac.uk}
]{\fnm{James} \snm{Deslauriers}}
\author[
 addressref={aff2},
 email={H.Dagdeviren@westminster.ac.uk}
]{\fnm{Huseyin} \snm{Dagdeviren}}
\author[
 addressref={aff2},
 email={resmiac@gmail.com}
]{\fnm{Resmi} \snm{Arjun}}
\author[
 addressref={aff2},
 email={H.Hamzeh@westminster.ac.uk}
]{\fnm{Hamed} \snm{Hamzeh}}


\address[id=aff1]{
  \orgdiv{School of Computing, Engineering \& the Built Environment},             
  \orgname{Edinburgh Napier University},          
  \city{Edinburgh},                              
  \cny{UK}                                    
}
\address[id=aff2]{%
  \orgdiv{School of Computer Science \& Engineering},
  \orgname{University of Westminster},
  \city{London},
  \cny{UK}
}
\address[id=aff3]{%
  \orgname{Institute for Computer Science and Control (SZTAKI), Eötvös Loránd Research Network (ELKH)},
  \city{Budapest},
  \cny{Hungary}
}
\address[id=aff4]{%
  \orgname{Department of Electronic and Electrical Engineering, University College London},
  \city{London},
  \cny{UK}
}





\begin{abstractbox}

\begin{abstract} 
IoT systems are becoming an essential part of our environment. 
Smart cities, smart manufacturing, augmented reality, and self-driving cars are just some examples of the wide range of domains, where the applicability of such systems has been increasing rapidly. These IoT use cases often require simultaneous access to geographically distributed arrays of sensors, and heterogeneous remote, local as well as multi-cloud computational resources. This gives birth to the extended Cloud-to-Things computing paradigm. The emergence of this new paradigm raised the quintessential need to extend the orchestration requirements i.e., the automated deployment and run-time management) of applications from the centralised cloud-only environment to the entire spectrum of resources in the Cloud-to-Things continuum. In order to cope with this requirement, in the last few years, there has been a lot of attention to the development of orchestration systems in both industry and academic environments. This paper is an attempt to gather the research conducted in the orchestration for the Cloud-to-Things continuum landscape and to propose a detailed taxonomy, which is then used to critically review the landscape of existing research work. We finally discuss the key challenges that require further attention and also present a conceptual framework based on the conducted analysis.
\end{abstract}




\end{abstractbox}
\end{fmbox}

\end{frontmatter}




\section{Introduction} \label{sec:introduction}
The advent of cloud computing has reshaped the way in which software is developed, deployed and used. Since its inception, the adoption of cloud services has continually increased. This is evident from the worldwide public cloud service revenue growth of 33 \%, from 266.4 billion dollars in 2020 to 354.6 billion dollars in 2022 \cite{gartner_2019}. This increased shift towards cloud computing is due to its inherent characteristics, such as on-demand provisioning, pay-as-you-go utility model and elasticity, which offer economic benefits as well as operational efficiencies to enterprises \cite{business_perspective}. 

In order to fully exploit the strength of cloud computing, effective and optimised usage of the associated computing resources is important. This is the responsibility of an orchestration system. More formally, an orchestration system automates the seamless delivery of applications over clouds, and guarantees various Quality of Service (QoS) goals, by handling the required complex tasks of resource selection, deployment, monitoring, and run-time control of the resources and applications \cite{Tomarchio2020}.   

In the last decade, cloud orchestration has become a mature research area and there emerged a large number of orchestration solutions. 
These include vendor-specific solutions, such as Amazon’s AWS Cloud-Formation \cite{AWSFormation2020}, OpenStack HEAT \cite{OpenStackHeat2020}, Microsoft Azure’s Resource Manager (ARM) templates \cite{ARMTemplate}, and Google’s Deployment Manager \cite{GoogleDeploymentManager}; some Open-source cloud agnostic initiatives, such as Kubernetes \cite{k8s}, Docker Swarm \cite{swarm}, Apache Brooklyn \cite{brooklyn}, Cloudify \cite{cloudify}, Cloudiator \cite{cloudiator}, Alien4Cloud \cite{alien}, MODAClouds \cite{modaclouds}  and MiCADO \cite{kiss2019micado}. 
The key purpose of all such tools is to improve resource utilisation and introduce a great deal of agility by making application development, deployment, execution and maintenance easier for cloud applications.

In recent years, the introduction of IoT has fuelled a new breed of applications, which in addition to cloud resources, also require IoT devices to capture and possibly process data from local environments. Such systems have a wide range of requirements in terms of low-latency analytics, data privacy and sensitivity, context awareness, time- and location- awareness, and simultaneous access to geographically distributed arrays of sensors, remote localised heterogeneous computational resources and to large-scale on-the-fly multi-cloud computational resources. A traditional cloud computing architecture is impractical, if not inadequate, to handle the aforementioned requirements. 
This gives rise to new computational paradigms such as fog computing, edge computing, and compute continuum.

The terms fog computing and edge computing are often used interchangeably to loosely refer to moving processing or computation away from the central cloud to nodes that are closer to endpoints at the network edge. Though they both aim to reduce the amount of data sent to the cloud in data-dense applications, there are subtle differences between the two. Fog computing is an intermediate layer between the cloud and edge that represents the nodes between the cloud to the IoT sensors and actuators, possibly spanning across multiple layers of the network topology. In contrast, in edge computing, the nodes where the computation takes place are normally very close to the IoT devices in terms of network proximity, often only one or a few hops away from the IoT devices, or even embedded within the connected device \cite{openfog}. 

The compute continuum---also known by other names such as cloud continuum, cloud-edge continuum, cloud-to-edge continuum, or cloud-to-things continuum---on the other hand, refers to the extension of cloud with energy-efficient and low-latency devices closer to the data sources located at the network edge~\cite{kimovski2021cloud}. More specifically, it extends the traditional Cloud towards multiple entities such as Fog, Edge, and IoT to provide different capabilities including analysis, processing, storage, and data generation~\cite{moreschini2022cloud}. Our adoption of the term Cloud-to-Things is to indicate the notion that the continuum connects cloud(s) and the IoT-connected devices (i.e., things)~\cite{moreschini2022cloud}, where we consider the ‘things’ mainly as a source of data that need to be processed in real-time using various layers of resources scattered across the continuum.

The emergence of these new paradigms raised the quintessential need to extend the orchestration requirements of applications from the centralised cloud-only environment to the entire spectrum of resources in the Cloud-to-Things continuum, as the existing cloud orchestration solutions are unable to address them. This mainly includes the application deployment and management to be performed in a more complex, heterogeneous and geographically distributed infrastructure, where resources are located across different layers of the continuum. More specifically, the following are some of the key challenges of orchestration in the Cloud-to-Things compute continuum~\cite{svorobej2020orchestration,bittencourt2018internet,hu2023architectural,ullah2021micado}:
\begin{enumerate}
    \item The Cloud-to-Things compute continuum is highly diverse, where resources are not only distributed across different layers of the spectrum but also heterogeneous having different architecture, operating systems, and computational capabilities. An orchestration system needs to provide seamless and simultaneous access to such a heterogeneous and decentralised resource landscape.
    \item The federated coordination across different administrative domains to facilitate end-to-end services across different cloud, fog and edge providers is challenging. An orchestration system needs standardised APIs and interfaces to achieve such coordination.
    \item A specific challenge to address in the case of edge nodes is to deal with volatility and mobility i.e.,  the nodes may shut down or lose connectivity or their locations may change. In such scenarios, the orchestration system needs to deal with resource fluctuation and changing environmental conditions. 
    \item Efficient monitoring mechanisms are required to collect the statuses of the workload and resource usage statistics across the entire spectrum of the continuum.
    \item The implementation of efficient run-time mechanisms that enforce policy-based deployment and run-time reconfiguration of target applications to ensure that the system meets the SLA goals specified in the form of contextual configurations in terms of resource discovery, optimal placement, optimise resource usage, efficient processing of data, and security aspects.
    \item The scale of the compute continuum can be massive, where resources can be gathered from different cloud and edge providers to fulfil the needs of target applications. An orchestration system needs to deal with the required level of scalability across the different administrative domains. 
    \item Lastly, an orchestration system is required to guarantee the security of the overall system against different attack scenarios while minimising the need for user-supplied configurations. 
    This is particularly challenging in the Cloud-to-Things continuum due to the heterogeneity of the resources involved and the possibility of their belonging to different administrative domains.
\end{enumerate}

To deal with the above-mentioned challenges, there has recently been a lot of attention, both in industry and academia, to the development of Cloud-to-Things Orchestration Solutions (CoTOS). This paper is an attempt to gather, analyse and synthesise the research work conducted in the field of orchestration systems for the Cloud-to-Things continuum. The key contributions of this paper are as follows:
\begin{enumerate}
	\item We identified a wide range of key characteristics in relation to the orchestration of IoT applications in the Cloud-to-Things computing continuum. These characteristics are the essential ingredients of, and therefore, important for the evaluation of CoTOS. Using these characteristics, a novel taxonomy of CoTOS is proposed, which is vital for the understanding and analysis of existing solutions. 
	\item We performed a thorough review covering a wide range of existing orchestration solutions from industry and 
academia that target the Cloud-to-Things continuum. The entire landscape of existing CoTOS is classified into different logical groups, and a detailed consolidated review and analysis of each group is performed in light of the proposed taxonomy.
	\item Based on the results obtained from the review, we identified and discussed the key issues and gaps in the existing landscape of orchestration solutions to highlight future research directions. 
	\item Lastly, we proposed a conceptual architecture of a novel and comprehensive orchestration framework as a reference to alleviate the identified gaps.
\end{enumerate}

The rest of this paper is structured as follows. Section~\ref{sec:relatedWork} discusses the existing related review papers to highlight the gaps and motivations in order to justify the need for conducting yet another review. Section~\ref{sec:taxonomy} presents our proposed taxonomy and explains each of the included characteristics. A thorough review of existing orchestration solutions, using the proposed taxonomy, is carried out in Section~\ref{sec:review}. Section~\ref{sec:discussion} further reflects on the summarised results to identify key issues and gaps from the review and to highlight research directions for the future. Section~\ref{sec:framework}, presents and discusses a conceptual framework that can be used as a reference for future implementations of orchestration solutions. Lastly, Section~\ref{sec:conclusion} concludes this paper.
\section{Related work} 
\label{sec:relatedWork}
This section discusses the most relevant review papers from the Cloud-to-Things orchestration domain, with a view to analyse their strengths and weaknesses and highlight how they differ from the review carried out in this work.

The most relevant studies related to the Cloud-to-Things orchestration include~\cite{Velasquez2018, lynn2020cloud, svorobej2020orchestration}. The authors in these papers have identified and discussed the target application scenarios and key challenges, in order to derive requirements that can be used for the design of a Cloud-to-Things orchestration solution. Based on these requirements, a detailed evaluation and analysis of some of the existing reference architectures and fog orchestration solutions have been provided. However, the list is not exhaustive and the authors, except in~\cite{svorobej2020orchestration}, have only covered a very small number of solutions. Furthermore, all these studies lack a detailed taxonomy. 

Similarly, papers~ \cite{Wen2017, Jiang2018, Comma-Di2018, velasquez2022resource} 
also identified and discussed the core issues and challenges related to the orchestration of IoT applications. The focus though in Karima, et al.~\cite{velasquez2022resource} is in the context of 5G (and beyond) networks. However, none of these papers provided a detailed taxonomy nor carried out a detailed review of existing orchestration solutions.

The authors in \cite{Nguyen2019} 
produced a systematic review of the deployment and orchestration approaches for the IoT. Their proposed taxonomy consists of the following three categories: (1) deployment and orchestration support, (2) specification, and (3) advance prospects (as defined by authors), 
i.e., monitoring, parameter adaptation, and trustworthiness features. As such, the authors used these very high-level characteristics only to classify the available solutions. In contrast, in this paper, we consider deployment and run-time management of IoT applications as the two essential key ingredients of an orchestration solution. Based on this notion, we used them as fundamental categories in our taxonomy. We further identified a large number of detailed, lower-level characteristics associated with these key ingredients to be part of the taxonomy. As a result, we review the existing approaches in light of these essential characteristics rather than the aforementioned high-level categories. Lastly, we also used additional aspects to classify the available approaches into different categories to provide a detailed comparative analysis and review of the overall spectrum of existing orchestration solutions.

The review by Wu in \cite{wu2020cloud} is mainly focused on 
the aspects related to architecture and AI-powered data processing techniques. The architecture was discussed in the context of an underlying communication infrastructure, such as the industrial network, mobile and vehicular networks; whereas, the data processing techniques are categorised and discussed based on the various functions from the orchestration viewpoint, such as Offloading, Placement, and Resource management. The scope of \cite{wu2020cloud} is on the different possible underlying architectures for the IoT ecosystem and the data processing techniques used by the applications. This paper, in contrast to \cite{wu2020cloud}, aims to perform a critical review of existing orchestration solutions.

Vaquero et al.~\cite{Vaquero2019} carried out an interesting and comprehensive review related to the challenges of next-generation service orchestration, where they focused on how the emergence of new technology trends such as Network Function Virtualisation (NFV), Software Defined Networking (SDN), Fog/Edge computing, and Serverless computing have changed requirements for the orchestration of microservices. Using the identified requirements, the authors further reviewed and discussed the state-of-the-art techniques by classifying them based on the implementation aspects, such as Machine learning techniques, P2P/Agent-based, Hierarchical and no orchestration. In contrast, our paper focused on the review of existing solutions with respect to the key functions of orchestration rather than their underlying implementation techniques.

The review in~\cite{bohm2022towards,bohm2022cloud} mainly focused on Kubernetes-based orchestration architectures that have been used within the context of the smart-city domain. Their key focus is on identifying the fundamental requirements for edge orchestration, analysis of existing Kubernetes-based architectures and in general the evaluation of Kubernetes as the suitable candidate for cloud-edge orchestration. The authors further reviewed and discussed the state-of-the-art Kubernetes architectures by classifying them mainly into three categories including frameworks that realise edge orchestration, solutions that implement custom modifications and extensions to Kubernetes, and solutions that only deal with edge layer using customised Kubernetes. All these categories in our paper are captured through only one category, titled Lower level (Please see Section~\ref{sec:review} for further details). Furthermore, we also include a range of other categories to cover the entire spectrum of cloud-to-edge orchestration solutions. Lastly, our paper provides a detailed taxonomy for cloud-to-edge orchestration, where the scope is also not limited to the smart city domain.

Fakude et al.~\cite{Fakude2019} and Šatkauskas et al.~\cite{vsatkauskas2020orchestration} have discussed fog/edge orchestration from the viewpoint of security in fog-enabled IoT-based computing environments. However, neither of these studies are detailed and only discusses a small number of existing works from the perspective of various security challenges. The focus in both studies is on the identification of security-oriented challenges related to fog orchestration. In the same realm, Al-Doghman~\cite{al2022ai} focused on highlighting the challenges of IoT management and secure decision-making at the edge for AI-based Microservices. All these studies, in comparison to our paper, do not provide a detailed taxonomy, formal classification and detailed analysis of existing orchestration solutions. Lastly, there is no discussion on identifying research gaps and future research directions.

Besides the above-mentioned studies, there are a number of considerably extensive review papers such as~\cite{hong2019resource, Hong2019, Tocze2018, Ghobaei-Arani2020,luo2021resource, duc2019machine} that have discussed the resource management related research works. The focus in these papers is on the classification of approaches that relate to resource management using different viewpoints. For example, the taxonomy proposed in~\cite{Ghobaei-Arani2020} relies on the classification based on the core functions such as application placement, resource scheduling, offloading\st{, etc.}. Similarly, Luo et al.~\cite{luo2021resource} focused on the core issues of computation offloading, resource allocation, and resource provisioning. On the other hand, the authors in~\cite{Tocze2018} use a set of criteria consisting of four points (type of resource, objective, resource location, and usage) to classify the available research work, where Duc et al~\cite{duc2019machine} reviewed machine learning techniques for resource provisioning in Edge-Cloud environment. All these papers provide a consolidated view of the available literature in the fog/edge computing area from a resource management point of view. In all these papers, there are either no or very limited attention provided to the automated orchestration of applications and resources.

In contrast to the above-mentioned related works, the research works in \cite{Tomarchio2020, Raj2018,Bellendorf2018,Ranjan2015,Weerasiri2017} focus specifically on orchestration. However, their scope is only limited to the cloud environment and does not cover the Cloud-to-Things ecosystem, 
as it is done in this paper. In terms of structure, our work extends and complements the taxonomies proposed in the aforementioned cloud orchestration review papers. 

To summarise, in contrast to the related works, our scope is on the overall key functions of orchestration rather than the underlying implementation techniques. As a result, this paper presents a detailed taxonomy of relevant characteristics, features, and dimensions related to the Cloud-to-Things orchestration. This taxonomy is further used as a unified framework to evaluate and perform a thorough analysis of existing orchestration solutions.
\section{Taxonomy}\label{sec:taxonomy}
\begin{figure*}[!ht]
\includegraphics[width=1.26\textwidth, angle=90]{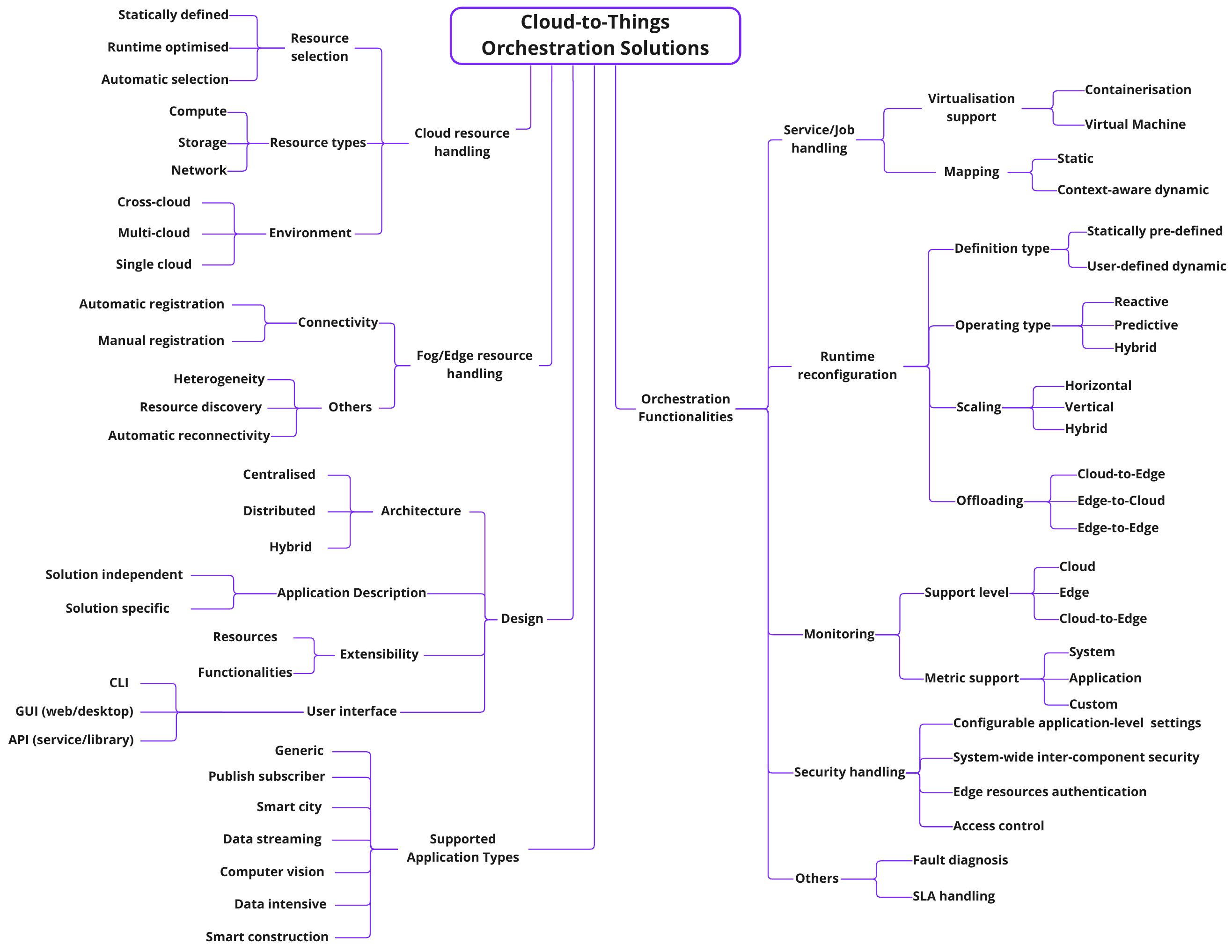}
\caption{Taxonomy of Cloud-to-Things orchestration}
\label{fig:taxonomy}
\end{figure*}

We identified a wide range of characteristics in relation to the orchestration of IoT applications in the Cloud-to-Things computing continuum using the various studies discussed in Section~\ref{sec:relatedWork}, literature review of target orchestration solutions, as well as our own experience of implementing an application-level cloud orchestration solution called MiCADO~\cite{kiss2019micado} and a CoTOS called MiCADO-Edge~\cite{ullah2021micado}. The identified characteristics of the taxonomy represent the essential ingredients of a CoTOS and therefore are important to be considered from an implementation viewpoint. Figure~\ref{fig:taxonomy} presents the proposed taxonomy, where the identified characteristics are structured and summarised under five main categories. 

The categorisation of attributes enabled us to demystify the concept and scope of orchestration for the continuum. However, all attributes are relevant to the entire orchestration solution and are not part of a specific type of resource environment, i.e., cloud or edge. The overall purpose of this taxonomy is to provide a unified framework whereby all candidate solutions can be objectively compared and evaluated. A brief description of each category and of its associated characteristics is provided in the following subsections. 
\subsection{Cloud resource handling}\label{sec:taxCloud}
This category groups together the characteristics related to the cloud infrastructure part of the orchestration system and includes the three aspects described below.
\paragraph{Environment.}
This attribute refers to the underlying support of a CoTOS for cloud environment(s) in terms of the ability to use or combine resources from different cloud providers. The possibilities include \textit{Single cloud},  where a CoTOS only supports a single specific cloud environment; \textit{Multi-cloud}, where a CoTOS facilitates the selection of suitable resources from multiple cloud environments, however, only one is utilised at a time; and \textit{Cross-cloud}, where 
multiple cloud environments are exploited simultaneously to allow the distribution of components belonging to the same application across different cloud providers. The use of multi and cross-cloud features is of particular importance as they help in optimising cost and performance by allowing the selection of suitable offers. This is also important to avoid vendor lock-in. Furthermore, it also helps in addressing privacy issues by allowing the use of specific cloud providers or private clouds for certain  application components~\cite{Tomarchio2020}.

\paragraph{Resource types.}
The support of a CoTOS in relation to the different types of resources that can be dynamically controlled by the orchestration solution. The choices include the three common utility services provided by cloud providers, i.e., compute, storage, and network.

\paragraph{Resource selection.}
The support of a CoTOS determines how the resources are selected during the deployment process. The available choices include \textit{Statically defined}, where specific instances of resources are statically assigned by the application owner at the time of deployment; \textit{Automatic selection}, where the application owner specifies the general resource characteristics and the CoTOS automatically selects the suitable instances at run-time, however, the selection does not change at run-time; \textit{Runtime optimised}, where the CoTOS automatically chooses the suitable resources from a diverse range of cloud vendors based on certain specific optimisation criteria, e.g., cost, locality.

\subsection{Fog/Edge resource handling}\label{sec:taxEdge}
This category groups together the characteristics related to the handling of resources from fog and edge (referred to as ``non-cloud'' collectively hereafter). More particularly, this covers the key aspects described in the following.
\paragraph{Heterogeneity}
The Cloud-to-Things computing continuum is highly heterogeneous, namely computational devices of different natures are usually required to support the requirements of an IoT application. This attribute will measure the support of CoTOS for device heterogeneity.

\paragraph{Connectivity.} In the Cloud-to-Things scenario, both dynamically created cloud resources (e.g., VMs) and non-cloud physical ones are available.
Therefore, the CoTOS has to provide a mechanism that enables the connectivity/registration of these non-cloud resources to a resource pool,
such that they can be utilised for the deployment as per the requirements of the IoT application. In this regard, the Connectivity attribute analyses the underlying support of a CoTOS that enables the connection of non-cloud resource elements to the pool of resources. This support can be further classified into two categories: i.e., \textit{Manual registration}, where the CoTOS facilitates users through some manual pre-defined procedure that allows the registration of non-cloud resources with the CoTOS prior to the deployment process; or  \textit{Automatic registration}, where the CoTOS provides an automatic procedure that allows the registration of non-cloud resources at run-time, even after the deployment process.

\paragraph{Automatic re-connectivity.} Non-cloud resources can be volatile in nature due to a number of reasons (e.g., low-powered computational devices, mobility, network connection), where they may lose connectivity to the rest of the system at different points in time. In this regard, the attribute refers to the ability (or not) of a CoTOS to support automatic re-connectivity of a non-cloud resource.

\paragraph{Resource discovery.} As resources in the Cloud-to-Things continuum are geographically distributed, it can be important for a CoTOS to support discovering all the available resources. Resource discovery refers to the ability of a CoTOS to support optimal re-configuration decisions by finding the most suitable resources based on certain contextual requirements.

\subsection{Orchestration functionalities}\label{sec:orcestrationFunc}
This category groups together the essential functions of a CoTOS.
\paragraph{Service/Job Handling.} This attribute determines the mechanism related to the deployment and management of services (or jobs in the case of batch-based applications). This can be further subdivided into the two aspects reported next.

\begin{enumerate}
    \item Virtualisation support: 
    this attribute can either refer to \textit{Virtual Machines (VMs)} to indicate that the CoTOS features the dynamic provision of VMs and the ability of direct deployment and management of application components on VMs without the use of containers; and \textit{Containerisation} to indicate that the CoTOS provides support for the deployment and management of application components through containers.
    
    \item Mapping: The mapping mechanism of application components to Cloud-to-Things resources can either be \textit{Static}, where the application owner statically configures application components to the available resources (or resource types); or \textit{Context-aware dynamic}, where the mapping is dynamically determined based on various user-defined contextual conditions associated with the application components and/or resources.
\end{enumerate} 

\paragraph{Run-time reconfiguration.} One of the key functions of any orchestration solution is its adaptability at run-time using reconfiguration of application components and associated resources according to the changing working conditions. We classify the \textit{Run-time reconfiguration} based on the list of attributes reported below.

    \begin{enumerate}
        \item Definition type: This represents the nature of the available reconfiguration functions. It can be one of the following two types: \textit{Statically pre-defined}, where a set of reconfiguration policies already exists and the application owners are restricted to provide threshold values based on some already established criteria; or \textit{User-defined dynamic}, where the application owners have the freedom to write their own policies based on available system and/or application-level metrics.
        \item Operating type: This represents the triggering behaviour of the reconfiguration operation. There are three possible types: \textit{Reactive}, where the reconfiguration is performed as a response to some changes;  \textit{Proactive}, where changes are anticipated and reconfiguration decisions are performed in advance; or  \textit{Hybrid}, meaning that the same solution consists of both reactive and proactive reconfiguration mechanisms.
        \item Scaling: This attribute represents the automated scaling ability of a CoTOS. It can be of the following three types: \textit{Horizontal}, where the number of additional resources is increased or decreased depending on the needs;  \textit{Vertical},  where the capacity of existing computational resources is increased or decreased; or \textit{Hybrid}, where the system supports both Horizontal and Vertical scaling.
        \item Offloading: This refers to the transfer of computational tasks from one execution device to another. For example, within the context of Cloud-to-Things computing, services are offloaded from cloud-to-edge devices due to latency sensitivity and/or geo-distributed requirements. Similarly, a service request can be offloaded from edge to cloud or another edge device if the existing device can not fulfil the required computational capacity. Consequently, three types of offloading generally occur. These are Cloud-to-Edge, Edge-to-Cloud, and Edge-to-Edge.
\end{enumerate}

\paragraph{Monitoring.} An orchestration system hugely relies on run-time monitoring through which information on the status of the system and of the application is gathered. The collected information is used to trigger run-time reconfiguration decisions in order to comply with system-stated objectives. Using this attribute, the monitoring support of a CoTOS can be evaluated in the following two aspects.

\begin{enumerate}
    \item Level of support: To identify whether monitoring of application components and resources is possible at each layer of the continuum or not. 
    \item Metrics support: The provided support of a CoTOS in gathering different types of metrics. The different types include \textit{system-level} (e.g., CPU/memory utilisation), \textit{application-level} (e.g., number of active HTTP requests), as well as the ability to define \textit{custom metrics} for collection (e.g., number of running jobs in a batch processing application).
\end{enumerate}

\paragraph{Security Handling.} A CoTOS is required to guarantee the security of the overall system against different attack scenarios while minimising the need for user-supplied configurations. Security handling in a CoTOS is a challenging task because an application in a Cloud-to-Things ecosystem typically runs on heterogeneous resources. Furthermore, these resources can contain low-powered devices that also operate in different administrative domains. Using this attribute, the support of CoTOS security handling will be evaluated according to the aspects reported below.

\begin{enumerate}
    \item Configurable application level security settings: The support of a CoTOS gives application owners the ability to define application level security settings (e.g., firewall setting, TLS/SSL certificate, ports configuration) in a configurable way.
    \item System-wide inter-component communication: The internal function of a CoTOS that enables secure communication amongst the different parts of a system, i.e., between system components that may operate in different VMs (and/or different layers).
    \item Edge resource authentication: An important function of a CoTOS is to facilitate the registration of non-cloud resources to the pool of resources that are then used for the deployment of application components. Such a registration process should be secure, where only authenticated non-cloud resources will be allowed to become part of the resources pool.
    \item Access control: This attribute will evaluate the support of CoTOS functionality in relation to the access control that includes aspects like secure access to the system resources.
\end{enumerate}

 \paragraph{Fault Diagnosis.} The support for the detection of system and/or application level faults at run-time, e.g., a fault in the cloud provider's system causing an unexpected termination of a VM, an unhandled run-time exception at application level forcing to stop a container, or a volatile edge node losing connection with the cluster.
\paragraph{Service Level Agreement (SLA) Handling.} This attribute will evaluate the support of a CoTOS in relation to the handling of SLA-related functionalities, such as specification, enforcement, and negotiation. 

\subsection{Design}\label{sec:taxDesign}
This category grouped together the design aspects of the orchestration solution. The following attributes are identified in what follows.

\paragraph{Architecture.} The employed architecture of a CoTOS influences how the overall system operates to perform the key orchestration functions, such as resource handling, application management, deployment and run-time reconfiguration decisions. This can be of one of the following three types: \textit{Centralised}, where a central entity, usually operating at the cloud layer, is responsible for all functions; \textit{Decentralised}, where multiple system entities are running at different Cloud-to-Things continuum layers and 
handle various orchestration functions accordingly.
\textit{Hybrid}, where a combination of the centralised and decentralised approaches are employed by the CoTOS. 

\paragraph{Extensibility.} The support provided by the design of a CoTOS for facilitating extension in terms of the addition of new resource providers, and the implementation of additional orchestration functions.

\paragraph{User Interface.} The ways users can interact with the CoTOS. The possible types include \textit{Graphical User Interface (GUI)}, \textit{Command Line Interface (CLI)}, and \textit{API-based Interfaces}.

\paragraph{Application description.} The orchestration solution usually provides application owners with a mechanism to provide the description of an application by expressing the specification of resources and components, the application topology, and any associated scaling and security policies. A number of well-known high-level description standards are available for this purpose, e.g., TOSCA~\cite{toscaweb}---an OASIS~\cite{oasis2020} standard for describing complex application topologies in the cloud. A standard TOSCA template in YAML defines the various components of a cloud application (software, storage, networks, virtual machines) as \textit{nodes}, which may have \textit{requirements} for, or share \textit{relationships} with, other nodes in the template. TOSCA also supports \textit{policies} for defining rules for scalability, monitoring, placement or security that will govern application behaviour at run-time.

The possible values are labelled as \textit{Solution independent} to represent that the provided mechanism is based on some standard and is independent of the underlying solution; or \textit{Solution specific} to represent that the provided mechanism is specifically designed for a given solution.

\subsection{Supported application types}\label{sec:supTargetApplication}
A CoTOS can be developed to target a specific application area. This attribute will evaluate a CoTOS in relation to its suitability with respect to 
particular application area/s. We treat this as an open-ended attribute, where specific application areas, such as \textit{Data streaming}, \textit{Computer vision}, \textit{Generic}, will be listed.
\section{Review of existing CoTOS} \label{sec:review}
This section presents a comprehensive review of existing CoTOS in light of the proposed taxonomy. The available landscape of orchestration solutions is very diverse, as subsets of solutions are quite different in nature from each other. Therefore, we first classified the landscape of existing solutions into different categories in order to reduce the overall complexity.  
The classification, on the one hand, allowed us to cover a representative solutions from each category. On the other hand, this enabled us to perform a detailed comparative analysis of solutions that are closely related to each other and to cluster the results of each category. The overall hierarchy of these categories can be seen in Figure~\ref{fig:classification} and their brief description is reported below.

\begin{figure}[!ht]
	\centering
	\begin{tikzpicture}[thick,scale=1.3, every node/.style={scale=0.6}, sibling distance=6.4em]
\node {\Large All solutions}
    child {node [draw = none] {\Large Lower level}
    edge from parent} 
    child {node [draw = none] {\Large Higher level}
    child {node [draw = none] {\Large Concept only}}
    child {node [draw = none] {\Large Production ready}
    child {node [draw = none] {\Large Research initiatives}}
    child {node [draw = none] {\Large Industry initiatives}}
    edge from parent}
    edge from parent};
\end{tikzpicture}
	\caption{Classification of existing orchestration solutions}
	\label{fig:classification}
\end{figure}

\begin{enumerate}
	\item Lower level: This category represents those solutions that act as middleware, lacking a high-level abstraction layer, and often requiring the knowledge and configuration of underlying low-level technical details in relation to setting up the infrastructure resources to be used for the application deployment. Furthermore, these solutions, also do not provide core orchestration functions such as deployment and reconfiguration based on user-provided dynamic criteria. Hence, these solutions cannot be directly considered as cloud-to-edge orchestrators. However, they are essential for higher-level application orchestrators to rely on as a middleware for the extension of orchestration capabilities to the edge.
	
	\item Higher level: This category represents a subset of solutions that hides the underlying complexity of resource settings and management using a high-level abstraction layer. Such an abstraction layer can be provided using a GUI and/or some standardised specification language, e.g., TOSCA. Such solutions may rely on the use of some lower-level solution (further discussed in Section~\ref{sec:lowerLevel}). From the application owners' viewpoint, the higher-level solutions are of particular interest. However, from the viewpoint of orchestration solution developers, the lower-level solutions are also relevant when analysing and selecting technologies that can support their higher-level orchestrator. Therefore, both are included in our analysis for completeness, and as a way to equally support application owners and orchestration solution developers. The higher level solutions are further classified  based on their existing status, i.e., \textit{Concept only}, which consists of those academic research proposals that only provide a conceptual framework and/or prototypical implementation only, and \textit{Production ready}, which consists solutions that provide a fully working implementation. The \textit{Production ready} solutions are further grouped into, Research initiatives that are developed as a result of some research projects, or Industry initiatives, where they are industry products and are commercially available.
\end{enumerate}

In the following subsections, we respectively review and analyse a subset of relevant orchestration solutions from each of the above categories. 
\subsection{Lower level solutions} \label{sec:lowerLevel}
All major public cloud providers such as Amazon, Microsoft, Google and Alibaba, provide middleware solutions that enable application developers to combine their edge resources and use them simultaneously with the respective cloud resources. Some examples of such solutions include AWS Greengrass~\cite{greengrass}, Azure IoT Edge~\cite{azureIoT}, Google Distributed Cloud Edge~\cite{googleEdge}, Alibaba Link IoT Edge~\cite{linkEdge}, IBM Edge Application Manager~\cite{ibmEdge}, and Akamai EdgeWorkers~\cite{akamiEdge}. These (and other similar) solutions have been developed with their respective cloud platforms in mind. Hence, they are not cross-platform solutions and cause a degree of vendor lock-in and therefore are not of particular interest for this review paper. 

There also exist a number of vendor-agnostic middleware solutions that fall into this category. A subset of such solutions are discussed in this section. The key factors that led to the inclusion of these solutions are 1) the availability of their implementation, 2) their implementations' being regularly maintained, and 3) the presence of available technical documentation and/or associated research papers. It is important to note that such middleware solutions can be used by higher-level application orchestrators to extend their capabilities to the edge. However, these solutions cannot be directly considered as cloud-to-edge orchestrators, as they lack some core essential features such as a high-level abstraction layer and dynamic deployment or reconfiguration based on user-provided criteria. Therefore, in this section, we only review these solutions and will not present their results in accordance with the taxonomy. The rest of this section discusses these solutions.

\textbf{Project EVE}~\cite{eve}, a Linux Foundation (LF) project, provides a flexible foundation for IoT edge deployments with a choice of any hardware, application, and cloud. EVE enables centralized scalable management of large volumes of edge compute nodes, where the orchestration of the underlying hardware and installed software is achieved through the open EVE API, which ensures consistency across diverse platforms. EVE is complementary to other LF Edge application frameworks including  Open Horizon \cite{openhorizon}, EdgeX Foundry \cite{edgexfoundry}, and Fledge \cite{fledge}. Open Horizon facilitates the management and deployment of workloads on edge devices from a management hub cluster. EdgeX Foundry provides an Edge IoT plug-and-play ecosystem with an aim to simplify and standardize edge computing architectures in the Industrial IoT market. Fledge is a Kubernetes-compatible container orchestrator for edge devices. Lastly, Fledge, in collaboration with the EVE system provides orchestration services and container run-time for Fledge-based applications.

\textbf{KubeEdge}~\cite{kubeedge} is another open-source initiative with a significant community behind it. KubeEdge extends native containerised application orchestration capabilities to non-cloud nodes at the edge of the network. It facilitates seamless and automatic configuration of edge nodes to make them part of a central Kubernetes cluster. KubeEdge empowers application developers to orchestrate apps, manage devices, and monitor application and system status at edge nodes, just like a normal Kubernetes cluster in the cloud. The components of KubeEdge facilitate the underlying infrastructure support for network, application deployment and synchronisation of metadata between cloud and edge. KubeEdge follows a centralised model. Hence, there is the risk of isolation for edge sites and therefore it is impossible to provision or reconfigure workloads hosted on non-cloud workers if the Kubernetes master node cannot be reached. 

In contrast to such a centralised approach, Kubefed \cite{kubefed} and Submariner \cite{submariner} follow a federated approach, where each edge site can continue to operate in case of network partitions. Such a federated approach offers the advantage of independent control over each edge site, in comparison to a single point of control as in the case of a centralised approach.

\textbf{Kubefed}, despite following a federated approach, still provides a unified way to manage the life cycle of a multi-cluster workload environment. Therefore, Kubefed can be considered a centralized server that distributes and propagates Kubernetes API objects to multiple clusters. It extends the Kubernetes API by leveraging the use of CustomResourceDefinitions, which is a mechanism to provide user-defined data types in Kubernetes. Overall, although Kubefed is able to provide cluster autonomy to a degree, it also presents several limitations to the edge use cases. For example, it does not implement any sort of knowledge or cooperation between the clusters themselves. Furthermore, the federation control plane, which is designed in a centralized manner, requires the re-creation of a lot of existing features at the federation level. 

\textbf{Submariner}, on the other hand, aims to solve the network connectivity between multiple Kubernetes instances. Unlike Kubefed, Submariner can expose Pods and Services from one cluster to another one without requiring a new API. Submariner relies on a few internal CustomResourceDefinitions (CRD) to make the inter-cluster communication possible. All the involved clusters synchronize their state to a shared cluster called Broker, which is responsible for the storage of all the CRD objects. Using this approach, Submariner succeeded in establishing interactions across the Services and Pods of independent clusters. However, the scalability and the robustness of sharing information are limited as most of the locally created objects, such as Deployments and Namespaces, remain local only.

\textbf{StarlingX}~\cite{starlingx}, similarly to KubeEdge, also extends native containerised application orchestration capabilities to the edges of the network, however, with two key distinctions: (1) StarlingX is specific to the use of OpenStack cloud, and (2) it forms independent edge  clouds in contrast to just connecting an edge node with a centralised cluster. The StarlingX solution has a central Kubernetes-based control centre called central cloud, with as many as required sub-clouds deployed on the edge nodes.
Using this model, StarlingX forms a federated architecture in a way, similar to that of KubeFed and Submariner. However, it still does not facilitate support for cross-cloud orchestration operations. Hence, all sub-clouds are independently controlled by their own controller. Unique features of StarlingX are its support of cluster management for services running on the HA (High Availability) master/control nodes cluster and recovery of services running on all nodes within the cluster.

\textbf{OpenIoTFog}~\cite{openiotfog} specifically focuses on the Industrial Internet of Things (IIoT) with two objectives: 1) to extend orchestration functions to the edge devices, and 2) to support the vision of Industry 4.0 by facilitating various related functions, such as real-time data aggregation, asset supervision, predictive maintenance, asset safety and the enablement of Digital Twins. The main aim of the OpenIoTFog is to provide software-based programmable logic controllers that can be dynamically updated and re-configured without production downtime. From a functional viewpoint, OpenIoTFog follows a similar model to that of KubeEdge, i.e., an agent component is required to be installed on an edge device, which makes that device part of a centralised cluster where dynamic policies, concerning the deployment of services on specific edge devices, can be applied. In addition, the agent component can also gather data from various sensors via industrial field bus systems and various (industrial) wireless technologies. It can also standardise, communicate and aggregate them through secured standard-compliant interfaces.

\textbf{Fornax}~\cite{fornax-project}, developed within the scope of an umbrella project called Centaurus~\cite{centaurus-project}, is an open-source edge computing framework for managing compute resources on edge environments. The key novel aspect of this project, amongst all the other ones described in this category, is its hierarchical topology that allows edge clusters to be formed and organised in a multi-layer tree-like structure. Hence, the infrastructure can be managed in N layers in comparison with the two-layered approach of KubeEdge and OpenIoTFog, or the federated approach adopted by other works including KubeFed, Submariner, and StarlingX.
\subsection{Higher level solutions} \label{sec:highLevel}
\subsubsection{Concept-only solutions} \label{sec:conceptOnly}
A large number of academic research papers are focused on the Cloud-to-Things orchestration aspects. It is not possible to cover all such papers individually and therefore, we shortlisted 10 papers from this category for review in detail, where other papers are briefly introduced. Amongst the 10 papers, half of them are the most highly cited papers of all time so far and the rest of them are all papers published in 2019 and onwards. It is important to note that these solutions are theoretical with no or just proof-of-concept implementation. The rest of this section discusses these solutions, where Table~\ref{table:conceptOnly} further presents a complete summary of the reviewed solutions in light of the attributes from the taxonomy.

\textbf{ENORM}~\cite{wang2017enorm}, a framework for edge node resource management, aimed to address the following three problems: (1) Edge node provisioning, (2) Workload deployment on edge with a focus on how to deploy and what services to deploy, and (3) resource management at the edge. ENORM follows a decentralised architecture, where edge nodes are responsible for their own resource management decisions. However, the overall architecture is static in nature, as all edge nodes are known in advance to the cloud servers' managers running in the cloud. The focus of ENORM is mostly on the operations of edge nodes, where it supports provisioning, monitoring, vertical scaling, and offloading applications. However, the details related to the cloud layer are not known, e.g., how the cloud server managers that are responsible for different applications are provisioned and maintained.

\textbf{Fernandez et al.}~ \cite{fernandez2019enabling} introduced slice orchestrator, which facilitates the automated orchestration of IoT services based on certain specific operational (and/or business) requirements over a set of shared infrastructures. Their idea is based on the 5G concept called network slice, which is an end-to-end logical network, capable of providing an agreed quality of service for a defined customer's purpose~\cite{gsma}. Based on this notion, an IoT slice would be a partition of the entire end-to-end IoT solution created to serve a specific (or a group of) customer(s). The job of the slice orchestrator is to establish network slices, set up edge and cloud tenants, and the deployment of IoT functions as per the specific requirements related to resources in terms of computing, storage, network, and target locations (e.g., edge and/or, cloud, and transport network). This solution followed a hybrid architecture, where a centralised slice orchestrator creates and manages slices but also relies on other domain-specific resource orchestrators (e.g., a different cloud orchestrator is responsible for a specific cloud environment) to perform key resource management functions such as resource selection and deployment. However, no details are provided regarding resource provisioning by domain orchestration, run-time reconfiguration aspects and the requirement specification that will be given as input to the system.

\textbf{Alam et al.}~\cite{alam2018orchestration} introduced a 3-layered reference architecture that makes use of Docker as the underlying orchestration tool for the automated deployment of microservices as containers. Their system follows a centralised model, where key functions like monitoring, adaptation, and orchestration take place at the cloud layer. Their Fog layer is mainly used as a gateway to mediate between the cloud and edge layers for system-specific operations (e.g., to update the status of connected edge devices) or application-specific operations (e.g., data transformation). This system is mostly suitable for publish-subscribe-based IoT applications. Similarly to Fernandez et al., \cite{fernandez2019enabling}, no details of various important functions,  such as device connectivity at the edge level, resource provisioning at different layers, and run-time reconfiguration, are provided. However, different to others, they  include a data mining component, which is responsible for erroneous behaviour detection such as responsiveness of deployed components, and edge devices' statuses. Hence, their system is adaptable in case of any failures.

\textbf{Santos et al.}~\cite{santos2017fog} focused on the optimal application placement problem in smart city applications while considering the reduction in network bandwidth usage and improved latency. Their proposal extends the ETSI NFV MANO architecture~\cite{etsiNFV} with additional functions of monitoring and data analysis. Their system follows a hybrid approach where management and decision-making related to the various functions happen at the cloud layer by cloud node (CN) and at the local layer by fog nodes (FNs). CN is responsible for the global view of the system including operations like coordination and control of FNs, global level data analysis and monitoring of the overall SLA. Each FN on the other hand has its own orchestrator and is responsible for autonomously managing its own local infrastructure, associated devices, and the life-cycle of microservices, as well as interfacing with the modules for resource discovery, system monitoring, data analysis, security, machine to machine communication, and decision making related to application life cycle and related policies. However, no details are provided in relation to these policies, their structure, or how they will be passed on to the system. This solution provides both GUI and API access to facilitate application owners managing and controlling FNs (and CN) independently and to perform manual updates if required. Lastly, a fog protocol based on the existing Open Shortest Path First (OSPF) routing protocol~\cite{ospf} has been proposed to enable and exchange communication between fog and edge layers. Details on edge device management, application description and run-time reconfiguration are missing.   

\textbf{Foggy}~\cite{foggy} framework, similarly to Santos et al.~\cite{santos2017fog}, aimed to minimise latency and perform optimal application placement. Foggy follows a centralised model. It consists of an orchestration server (OS)---a central entity responsible for deployment and resource management decisions---and an orchestration client (OC)---running on each computational resource and is responsible for enforcing deployment decisions.
Overall, Foggy offers the following unique characteristics in contrast to  other solutions discussed in this category: 1) To facilitate an automated build, a direct integration of a version control system (such as Github) and continuous integration process as part of their system architecture; 2) A pluggable policy-driven deployment planner that dynamically identifies suitable resources based on user provided requirements; 3) A JSON based container specification to facilitate application owners to provide service requirements using qualitative constructs such as Low, Medium, and High. However, it is not clear how these qualitative specifications for different aspects, such as computation and latency, are mapped within the system. Similarly to others, Foggy also does not cover details related to edge device registration, standardised application description and run-time reconfiguration.

\textbf{Castellano et al.}~\cite{castellano2019service} solution follows a distributed approach where a dedicated instance of a service-defined orchestrator (SDO) is initiated every time a new application is deployed. The input to the system is an application deployment request that mainly consists of a list of components, their topology and a set of declarative statements to form the Orchestration Behaviour Model (OBM) that drives the orchestration functions. The OBM features aspects, such as infrastructure and/or application state, required objectives to be optimised, the events and the corresponding actions to be performed. Using the OBM, every SDO instance aims to make optimal decisions with respect to the managed application. However, this also raises the resource allocation issue for different instances of SDOs at the shared infrastructure level when resources are limited. To cope with this, Castellano et al.~\cite{castellano2019service} introduced Dragon---an additional component responsible for the optimal partitioning of the underlying shared resources across different SDOs. Using Dragon, the SDO can decide to terminate an application component if it cannot allocate the required resources to that particular component. Their proposed declarative statements-based application description approach, however, is specific to this solution only and does not follow a standardised approach.

\textbf{HYDRA}~\cite{jimenez2020hydra}, similar to Castellano et al.~\cite{castellano2019service} also follows a decentralised architecture, where a set of distributed nodes without the presence of a centralised entity are responsible for performing the orchestration functions of one application. HYDRA actually builds a peer-to-peer (P2P) overlay network of computational nodes, where every node serves both as an orchestrator as well as a computational resource---responsible for running the application micro-services. HYDRA supports both location-agnostic as well as location-aware application deployment with a primary focus on the overall scalability and resilience aspects of the underlying resource infrastructure through its decentralised architecture. This has been achieved through the adoption of a dynamic partitioning scheme, where orchestrator nodes operate independently to control the needs on per application basis.

\textbf{Caravela}~\cite{pires2021distributed} follows a similar decentralised model, where all key aspects such as the overall architecture, resource discovery and scheduling are also based on the concept of a P2P overlay network. However, different from HYDRA, it follows a market-oriented approach, where volunteer resources can join the ecosystem and get rewarded for their services. Caravela dynamically builds edge cloud from the volunteer resources that are further used to deploy applications using Docker containers. Caravela's scope, however, is only limited to non-cloud layers and does not include resource provisioning from the cloud. 

\textbf{Mathias et al.}~\cite{de2017service} solution consists of a Fog Orchestrator (FO)---a central entity responsible for maintaining a resource catalogue of fog nodes, overall service management, global level monitoring, and orchestration---and an agent component called Fog Orchestrator Agent (FOA) that runs on every fog node and is responsible for activities such as management of connected edge devices, security and monitoring. The working mechanism of this solution suggests that FO composes a TOSCA-based orchestration template using information obtained from a resource catalogue and monitoring components. This template is further used for deployment and run-time management. Such usage of TOSCA for expressing orchestration strategies is common and has been used by many solutions such as~\cite{ullah2021micado, tsagkaropoulos2021extending, cloudify, sodalite}. However, in this case, the TOSCA template is dynamically generated by the system and, therefore, it is not clear what the initial input to the system is. A unique prospect of this solution, in contrast to others discussed in this category, is that the FOA can also act as FO, if the connection is lost between them. However, this behaviour is static and the specific fog node has to specify this at the time of joining. Furthermore, the scope of the overall solution only includes the non-cloud layers. 

\textbf{Hetero-Edge}~\cite{zhang2019hetero} follows a similar concept to that of Mathias et al.~\cite{de2017service}, where a central entity has been used to handle the orchestration functions at the non-cloud (edge) layer only. The solution, however, is specific to computer vision applications and relies on the use of Apache Storm (or something similar, such as Apache Flink). Hetero-Edge breaks down an application into smaller Apache Storm tasks and then efficiently maps them onto the connected edge nodes with the objective of minimising the overall end-to-end latency. The specification of tasks is provided through a directed acyclic graph, where the mapping is performed using their custom-developed task scheduler that takes into account the estimated performance and resource demands of tasks. The solutions proposed by Donassolo et al.~\cite{donassolo2019fog} also follow a similar model, which supports orchestration at the non-cloud layers, however, with a particular focus on optimising the provisioning cost of IoT applications.

Some other more recent notable contributions include GeneSIS~\cite{ferry2019genesis}, which proposed a model-driven approach to automate the deployment of different kinds of deployable artefacts including binary, ThingML-based~\cite{morin2016generative}, and
container; ECCO~\cite{cozzolino2020ecco}, which proposed an orchestration framework for enabling the collective use of edge-cloud resources for
road context assessment; KubeHICE~\cite{yang2021kubehice}, which took on the challenge of addressing hardware heterogeneity by automatically matching the right computational device that is compatible with the instruction set architecture (ISA) supported by the containerized application; and Gand et al.~\cite{gand2020fuzzy} and Sonmez et al.~\cite{sonmez2019fuzzy}, which focused on the presence and importance of uncertainty in the cloud-to edge environment and therefore adopted a fuzzy logic-based approach for workload deployment.

In addition to the above-mentioned solutions, there are also some research works that did not directly cover the core orchestration functions, however, they emphasised the importance of other related aspects. For example, the authors in \cite{pahl2018architecture, el2018trustworthy} introduced the notion of trusted orchestration, where the proposed approach aimed at identifying and tracking orchestration activities to improve trust across the involved actors of the system. Similarly, Kochovski et al.~\cite{kochovski2020smart} proposed a smart contract (SC) based architecture for SLA management and verification amongst relevant entities and actors of a decentralised environment. More recent works on the DRL-based advanced techniques for dynamic load balancing~\cite{liu2021deep}
and network dynamic clustering~\cite{liu2019deep} in edge computing  focused on the overall optimisation of cloud-to-edge system. Such solutions can be integrated into distributed orchestration solutions to support self-organisation and optimisation behaviours. Lastly, with the growing popularity of Deep Learning (DL) applications, there is also an increasing interest in proposing  resource management solutions that are specifically tailored to DL applications. For example, FlowCon~\cite{mao2022elastic} monitors the execution of DL jobs at run-time to make informed resource allocation and placement decisions. Similarly, SpeCon~\cite{mao2021speculative} is a container scheduler that aims to optimise resource usage and improve the performance of DL training jobs, whereas DQoES~\cite{mao2022differentiate} aims at dynamically adjusting cloud resources to meet the target Quality of Experience (QoE) specified by the clients. The scope of all the aforementioned DL-tailored solutions, however, only includes the cloud layer. The details of these papers do not directly fall within the scope of the proposed taxonomy and therefore have not been included here in larger details.
\begin{table*}[tp]
\renewcommand{\arraystretch}{1.4}
\scriptsize
\caption{Comparative summary of the concept-only orchestration solutions}
\begin{tabular} { | c | p{0.17\textwidth} | p{0.12\textwidth} | p{0.17\textwidth} | c | c | c | c | c | c | c | c | c | c |} 
 \hline
  \multicolumn{4}{|c|}{Attributes} & \rotatebox{90}{ENORM \cite{wang2017enorm}} & \rotatebox{90}{Fernandez et al. \cite{fernandez2019enabling} } & \rotatebox{90}{Alam et al.~\cite{alam2018orchestration} } & \rotatebox{90}{Santos et al.~\cite{santos2017fog} } & \rotatebox{90}{Foggy \cite{foggy} } & \rotatebox{90}{Castellano et al.~\cite{castellano2019service} } & \rotatebox{90}{HYDRA~\cite{jimenez2020hydra}} & \rotatebox{90}{Caravela~\cite{pires2021distributed} } & \rotatebox{90}{Mathias et al.~\cite{de2017service} } & \rotatebox{90}{Hetero-Edge~\cite{zhang2019hetero} }  \\ \hline

\multicolumn{2}{|c|}{\multirow{9}{*}{ Cloud resource handling} } & \multirow{3}{*}{Environment} & Single cloud& \checkmark &   & \checkmark &  & \checkmark  & \checkmark &  &  &  & \\ \cline{4-14}
 
 \multicolumn{2}{|c|}{} &  & Multi-cloud &  & \checkmark  &  & \checkmark &   &  & \checkmark & &  &  \\ \cline{4-14}
   
 \multicolumn{2}{|c|}{} &  & Cross-cloud &  &   &  &  &   &  & & &  & \\ \cline{3-14}
    
 \multicolumn{2}{|c|}{} & \multirow{3}{*}{Resource types} & Compute & \checkmark & \checkmark  & \checkmark & \checkmark & \checkmark  & \checkmark & \checkmark & \checkmark &  & \\ \cline{4-14}
 
 \multicolumn{2}{|c|}{} &  & Storage &  & \checkmark  &  &  &   & \checkmark & & \checkmark &  & \\ \cline{4-14}
 
 \multicolumn{2}{|c|}{} &  & Network &  & \checkmark  &  &  &   & \checkmark & & \checkmark &  &  \\ \cline{3-14}
 
 \multicolumn{2}{|c|}{} & \multirow{3}{*}{Resource selection} & Statically defined & \checkmark & \checkmark  & \checkmark &  & \checkmark  &  & &  &  &  \\ \cline{4-14}
 
 \multicolumn{2}{|c|}{} &  & Automatic selection &  &   &  &  &  \checkmark & \checkmark & \checkmark & \checkmark &  & \\ \cline{4-14}
 
 \multicolumn{2}{|c|}{} &  & Run-time optimised &  &   &  &  & \checkmark  & \checkmark & & &  &  \\ \hline
 
 
 \multicolumn{2}{|c|}{\multirow{5}{*}{Fog/Edge resource handling} }  & \multirow{2}{*}{Connectivity} & Manual registration & \checkmark & \checkmark  & \checkmark &  & \checkmark  & \checkmark & \checkmark &  &  & \checkmark \\ \cline{4-14}
 
 \multicolumn{2}{|c|}{} &  & Automatic registration &  &   &  & \checkmark &   & & & \checkmark & \checkmark &  \\ \cline{3-14}
 
 \multicolumn{2}{|c|}{} & \multirow{3}{*}{Others} & Heterogeneity & \checkmark & \checkmark  & \checkmark & \checkmark & \checkmark  &  & \checkmark & \checkmark & \checkmark & \checkmark  \\ \cline{4-14}
 
 \multicolumn{2}{|c|}{} &  &  Auto reconnectivity &  &   &  &  &   &  & & \checkmark &  &  \\ \cline{4-14}
 
 \multicolumn{2}{|c|}{} &  & Resource discovery &  &   &  & \checkmark & \checkmark  &  & \checkmark & \checkmark & \checkmark &  \\ \hline
 
 \multirow{28}{*}{\rotatebox{90}{Orchestration functionalities}}  &  \multirow{4}{*}{Service/Job Handling}  & \multirow{2}{*}{Virt support}  & VM & \checkmark & \checkmark &  & \checkmark & \checkmark  & \checkmark & &  &  &  \\ \cline{4-14}
 
  & & & Containerisation & \checkmark & \checkmark  & \checkmark & \checkmark & \checkmark  & \checkmark & \checkmark & \checkmark & \checkmark & \checkmark  \\ \cline{3-14}
  
   & & \multirow{2}{*}{Mapping} & Static& \checkmark & \checkmark  & \checkmark & \checkmark & \checkmark  &  & & \checkmark & \checkmark &   \\ \cline{4-14}
 
   & & & Context aware&  &   &  &  & \checkmark  & \checkmark & \checkmark & &  & \checkmark \\ \cline{2-14}
 
 & \multirow{11}{*}{Run-time reconfiguration} & \multirow{2}{*}{Definition type} & Statically pre-defined& \checkmark &   &  &  &   &  & & & \checkmark &   \\ \cline{4-14}
 
 & & & User-defined dynamic&  &   &  &  &   & \checkmark & & &  &   \\ \cline{3-14}
 
 & & \multirow{3}{*}{Operating type} & Reactive& \checkmark &   &  &  &  \checkmark & \checkmark & \checkmark &  &  &   \\ \cline{4-14}
 
 & & & Proactive&  &   &  & \checkmark &   &  & & &  &   \\ \cline{4-14}
 
 & & & Hybrid&  &   &  &  &   &  & & &  &   \\ \cline{3-14}
 
 & & \multirow{3}{*}{Scaling} & Horizontal&  &   &  &  &   &  & \checkmark & &  &   \\ \cline{4-14}
 
 & & & Vertical& \checkmark &   &  &  &   &  & & &  &   \\ \cline{4-14}
 
 & & & Hybrid &  &   &  &  &   & \checkmark & & & \checkmark &   \\ \cline{3-14}
 
 & & \multirow{3}{*}{Offloading}& Cloud-to-Edge& \checkmark &   & \checkmark & \checkmark &   & & &  &  &  \\ \cline{4-14}
 
 & & & Edge-to-Cloud& \checkmark &   & \checkmark &  &   &  & & &  &  \\ \cline{4-14}
 
 & & & Edge-to-Edge&  &   &  & \checkmark &   &  & & &  &  \\ \cline{2-14}
 
 & \multirow{6}{*}{Monitoring} & \multirow{3}{*}{Support level} & Cloud&  &   & \checkmark &  &   & & & &  &  \\ \cline{4-14}
 
 & & & Edge& \checkmark &   &  &  &   &  & & \checkmark & \checkmark & \checkmark \\ \cline{4-14}
 
 & & & Cloud-to-Edge&  &   & \checkmark &  & \checkmark  & \checkmark & \checkmark & &  &   \\ \cline{3-14}
 
 & & \multirow{3}{*}{Metrics support} & System&  &   &  & \checkmark & \checkmark  & \checkmark & \checkmark & \checkmark & \checkmark & \checkmark  \\ \cline{4-14}
 
 & & & Application& \checkmark &   & \checkmark &  & \checkmark  & \checkmark & & &  & \checkmark  \\ \cline{4-14}
 
 & & & Custom&  &   &  &  &   &  &  & &  &   \\ \cline{2-14}
 
 & \multicolumn{2}{c|}{\multirow{4}{*}{Security handling} } & Configurable app level& \checkmark &   &  &  & \checkmark  &  & & &  &  \\ \cline{4-14}
 
 & \multicolumn{2}{c|}{}  & Sys wide inter-comp&  &   &  & \checkmark &   & & & &  &   \\ \cline{4-14}
 
 & \multicolumn{2}{c|}{}  & Edge authentication&  &   &  &  &   & & & & \checkmark &   \\ \cline{4-14}
 
 & \multicolumn{2}{c|}{} & Access control&  &   &  &  &   & & \checkmark & &  &   \\ \cline{2-14}
 
 & \multicolumn{2}{c|}{\multirow{2}{*}{Others}} & Fault diagnosis&  &   & \checkmark & \checkmark &   & & &  &  &   \\ \cline{4-14}
 
 & \multicolumn{2}{c|}{} & SLA Handling& $\sim$ &   &  &  &   & $\sim$ & & &  &   \\ \hline
 
 \multicolumn{2}{|c|}{\multirow{10}{*}{Design} } & \multirow{3}{*}{Architecture} & Centralised&  &   & \checkmark &  & \checkmark  &  & & &  \checkmark & \checkmark \\ \cline{4-14}
 
 \multicolumn{2}{|c|}{} &  & Decentralised & \checkmark &   &  &  &   & \checkmark & \checkmark & \checkmark &  &  \\ \cline{4-14}
 
 \multicolumn{2}{|c|}{} &  & Hybrid &  & \checkmark  &  & \checkmark &   &  & & &  &  \\ \cline{3-14}
    
 \multicolumn{2}{|c|}{} & \multirow{2}{*}{App description} & Solution independent &  &   &  &  &   &  & & &  \checkmark & \checkmark \\ \cline{4-14}
 
 \multicolumn{2}{|c|}{} &  & Solution specific &  &   &  &  & \checkmark  & \checkmark & \checkmark & \checkmark &  &  \\ \cline{3-14}

 \multicolumn{2}{|c|}{} & \multirow{2}{*}{Extensibility} & Resources &  &   &  &  &   & & &  &  &  \\ \cline{4-14}
 
 \multicolumn{2}{|c|}{} &  & Functionalities &  &   &  &  & \checkmark  & & \checkmark & \checkmark &  &  \\ \cline{3-14}
 
 \multicolumn{2}{|c|}{} & \multirow{3}{*}{User interface} & GUI (Web/Desktop) &  &   &  & \checkmark &   &  & & &  &  \\ \cline{4-14}
 
 \multicolumn{2}{|c|}{} &  & CLI & \checkmark & \checkmark  &  &  & \checkmark  & \checkmark & & \checkmark & \checkmark & \checkmark \\ \cline{4-14}
 
 \multicolumn{2}{|c|}{} &  & API (Service/Library) &  &   &  & \checkmark &   & & & \checkmark &  &  \\ \hline
 

 \multicolumn{3}{|c}{} & Supported App types & G & G  & PS & SC & G  & DS & G & G & G & CV  \\ \hline
 
\end{tabular}
 {\raggedright [Supported = \checkmark, partially supported = $\sim$] Supported App types $\Rightarrow$ Publish Subscriber \textbf{(PS)}, Smart City \textbf{(SC)}, Data Streaming \textbf{(DS)}, Computer Vision \textbf{CV}, Generic \textbf{(G)}  \par}
\label{table:conceptOnly}
\end{table*}
\subsubsection{Production ready solutions} \label{sec:productionReady}
\paragraph{\textbf{Research initiatives}}
In the last few years, a number of EU-funded research projects focused on developing cloud-to-edge solutions. The selection of this subset was made considering three key aspects: 1) whether their core functionality was related to the cloud-to-edge orchestration, 2) whether their implementation was available, and 3) whether there were any publications associated with the solution. The rest of this section discusses these solutions, and Table~\ref{table:research_oriented} further presents a complete summary of the reviewed solutions in light of the attributes from the taxonomy.

\textbf{SODALITE@RT}~\cite{sodalite} supports the deployment and management of applications across a cloud-to-edge infrastructure in a portable manner. The term ``portable'' is based on their use of TOSCA as the deployment model to represent application components and resources; and the use of the Infrastructure as Code (IaC) concept~\cite{morris2016infrastructure} to implement the life-cycle operations of components, for which they utilised Ansible ~\cite{ansible2020}. SODALITE@RT follows a centralised model, where a central component called meta-orchestrator receives TOSCA-based deployment models and Ansible implementation scripts to set up the resources and to perform the deployment. The Ansible scripts are cloud provider specific that the orchestrator pulls from an IaC repository. Such an approach enables custom implementation, however, also burdens application developers with the production of Ansible scripts in comparison with other TOSCA-based solutions such as~\cite{ullah2021micado, tsagkaropoulos2021extending}, where the TOSCA model is the only input. The Ansible scripts take care of the cloud resources handling, where the edge resources are handled as part of a Kubernetes cluster. However, details on edge cluster formation are not provided and therefore it is not clear whether a meta-orchestrator creates the edge cluster or it must exist prior to the deployment process. SODALITE@RT also provides an event-condition-action-based policy language to support custom redeployment policies. Furthermore, it also supports access control and a mechanism for secure storage of application secrets. However, no mechanism for application-level security configurations is provided.

\textbf{Capillary}~\cite{capillary} focused on the use of a custom-built monitoring system to measure QoS parameters and Offloading across different resource layers based on various user-defined characteristics, including geographic positioning. The offloading decisions follow an ``offload to next immediate layer'' model (e.g., edge to fog or fog to cloud) that resembles the capillary fluid movement, hence the name Capillary. It follows a centralised approach, where a central entity called a Capillary container orchestrator performs the deployment and offloading operations. The input to the system is a TOSCA deployment model that includes various details, such as resource capacity requirements for services, zone details, and constraints on QoS thresholds that are used for reconfiguration purposes. At run-time, the monitoring system raises alarms based on the developer-provided thresholds. As a result, a sub-component of the orchestrator, similar to SODALITE@RT~\cite{sodalite}, takes an offloading decision, changes the TOSCA model and triggers re-deployment. For resource handling, the cloud resources are dynamically provisioned by the orchestrator based on the user-provided minimum requirements for the service. However, no details on the provisioning of the fog and edge infrastructure are provided.

\textbf{MiCADO-Edge}~\cite{ullah2021micado} is also a centralised solution, where a central entity called MiCADO-Master is responsible for the automated deployment and management of a microservices-based application across the cloud-to-edge continuum using a single TOSCA based deployment model. This model consists of details related to computational resources, component specification, application topology, service placement mapping, user-defined scaling policies and any application-specific security settings. The key focus in MiCADO-Edge is on generalising the resources across the different layers of the continuum by facilitating a mechanism to allow the resources from fog and edge layers (referred to as non-cloud resources) to join a centralised cluster prior to the application deployment process. Once they become part of the MiCADO cluster, developers can reference them in the TOSCA-based deployment model to define placement and reconfiguration policies. Furthermore, MiCADO-Edge empowers application developers to write custom dynamic scaling policies based on a wide range of application and system metrics. MiCADO-Edge, however, currently lacks support for context-based placement of application services and developers are required to provide static mapping between services and resources. 

\textbf{PrEstoCloud}~\cite{verginadis2021prestocloud, verginadis2017prestocloud} follows a similar model of a TOSCA-based orchestration solution. However, it provides an optimisation step before deployment. This step consists of receiving TOSCA in a high-level form (type level TOSCA model as they referred to it), which also contains optimisation criteria independent from the underlying infrastructure resources. Based on the provided criteria, the solution automatically produces a more specific instance-level TOSCA deployment model containing the specific resources across the infrastructure that are to be used for application deployment. Hence, it provides an optimised placement mechanism. Furthermore, PrEstoCloud also focused on facilitating predictive reconfiguration based on the changing data stream conditions considering data-intensive applications.       

\textbf{mF2C}~\cite{masip2021managing} adopted an N-layered approach to utilise the available resources in the continuum from edge (Layer-N) to the cloud (Layer-0), in contrast to the two-layered (i.e., cloud and non-cloud) approach followed by MiCADO-Edge~\cite{ullah2021micado} and the typical three-layered approach as followed in~\cite{capillary}. Their proposed solution is decentralised, where deployed mF2C agents, coordinate with each other to find suitable resources, closer to the edge, for the execution of application services. The input to the system, e.g., a service execution request is received by the mF2C agent at the lower layer. The receiving agent, in coordination with other agents at the same layer, aims to execute the service request if the required resource specification can be fulfilled. Otherwise, the request is further passed on to the mF2C agents at the upper layer. The service execution request is in JSON format that includes the required resource specification used by the mF2C agents to make deployment decisions. In terms of resource handling, the mF2C architecture supports the automatic discovery of other mF2C agents, the dynamic formation of clusters, and also reconfiguration in case of device mobility prospects. However, it does not address aspects like scaling, offloading, dynamic provisioning of resources, and configurable policies.

\textbf{DECENTER}~\cite{kochovski2021building} is specifically developed for transforming construction sites into smart and safe environments. Hence, this solution facilitates methods that are specifically tailored to the problems related to construction processes. The unique feature of DECENTER, amongst other solutions in this category, is the Blockchain-based resource brokerage mechanism, which facilitates the trusted brokerage and negotiation of computational resources that can be used for deployment. Furthermore, all transactions of the system are traceable and can be formally verified. Hence, improving the trust and transparency of the overall system. DECENTER follows a centralised architecture, where four key components of the system including Application composer, QoS-aware decision maker, Monitoring system, and Orchestrator are responsible for performing the key orchestration functions. It also facilitates users with a GUI interface to select the services they want to use and define their QoS objectives. These details, along with the monitoring data, are used by the QoS-aware decision maker to perform deployment decisions that are forwarded to the orchestrator. DECENTER also supports automatic redeployment, when the system encounters violation of QoS specifications. However, it lacks functions like dynamic auto-scaling and offloading. 

\textbf{Pledger}~\cite{pledger-project} also makes use of Blockchain to improve trust, secure communication and to enable ad-hoc networks between the resources of non-cloud layers to collaborate with each other for the execution of a specific application. Although Pledger's overall architecture follows a centralised model, its implementation does not comply with a traditional adapter-based interaction model between different parts of the system. Pledger provides different tool-kits for resource providers to integrate their resources into the Pledger ecosystem and for the application owners to perform mapping of their applications on specific resources, which is further assessed and reconfigured by the core Pledger service to ensure optimised use.        

\textbf{Rainbow}~\cite{rainbow-project} particularly focused on the issue of lack of handling concerning the fog-specific constraints related to the deployed services. For this purpose, their proposed solution consists of a high-level abstraction mechanism, where application topology and the related constraints on services are described through a graph. The Rainbow orchestration system accepts the graph as input and deals with the optimised placement of the services and the execution thereafter. The orchestration system follows a decentralised model, where different components of the system may run on the different computational nodes that are part of the Rainbow ecosystem. To address the various challenges of the fog environment (such as low-powered devices, intermittent connectivity, and the interactions of sub-components), the system follows a publish-subscribe mechanism where a component called Orchestrator Repository maintains the states of the system and its sub-components. The Rainbow platform facilitates the dynamic registration of edge devices and their reconfiguration as per defined service level objectives (SLO) violations. However, its scope is only limited to fog/edge resources and lacks the dynamic provisioning of cloud resources. 
\begin{table*}[tp]
\renewcommand{\arraystretch}{1.4}
\scriptsize
\caption{Comparative summary of research projects based orchestration solutions}
\begin{tabular} { | c | p{0.17\textwidth} | p{0.14\textwidth} | p{0.17\textwidth} | c | c | c | c | c | c | c | c | c |} 
 \hline
  \multicolumn{4}{|c|}{Attributes} & \rotatebox{90}{SODALITE@RE~\cite{sodalite} } &
  \rotatebox{90}{Capillary~\cite{capillary} } & 
  \rotatebox{90}{mF2C~\cite{masip2021managing} } &
  \rotatebox{90}{MiCADO-Edge~\cite{ullah2021micado}} &
  \rotatebox{90}{PrEstoCloud~\cite{verginadis2021prestocloud, verginadis2017prestocloud} } & 
  \rotatebox{90}{DECENTER~\cite{kochovski2021building} } & 
  \rotatebox{90}{Rainbow~\cite{rainbow-project} } & 
  \rotatebox{90}{Pledger~\cite{pledger-project}} &
  \rotatebox{90}{Slack4things~\cite{slack4things, merlino2019enabling}} 
  
  \\ \hline

\multicolumn{2}{|c|}{\multirow{9}{*}{ Cloud resource handling} } & \multirow{3}{*}{Environment} & Single cloud&  & \checkmark  &  &  &   &  &  & & \checkmark \\ \cline{4-13}
 
 \multicolumn{2}{|c|}{} &  & Multi-cloud &  &   & \checkmark & \checkmark &  &  &  & \checkmark & \\ \cline{4-13}
   
 \multicolumn{2}{|c|}{} &  & Cross-cloud & \checkmark &   &  &  & \checkmark  & \checkmark &  &  & \\ \cline{3-13}
    
 \multicolumn{2}{|c|}{} & \multirow{3}{*}{Resource types} & Compute & \checkmark & \checkmark  & \checkmark & \checkmark & \checkmark  & \checkmark & \checkmark & \checkmark & \checkmark \\ \cline{4-13}
 
 \multicolumn{2}{|c|}{} &  & Storage &  &   &  &  &   &  &  & & \checkmark  \\ \cline{4-13}
 
 \multicolumn{2}{|c|}{} &  & Network &  &   &  &  &   &  &  & &  \\ \cline{3-13}
 
 \multicolumn{2}{|c|}{} & \multirow{3}{*}{Resource selection} & Statically defined & \checkmark &   &  & \checkmark &   &  &  & \checkmark & \checkmark  \\ \cline{4-13}
 
 \multicolumn{2}{|c|}{} &  & Automatic selection &  & \checkmark  & \checkmark &  & \checkmark  & \checkmark & \checkmark &   & \\ \cline{4-13}
 
 \multicolumn{2}{|c|}{} &  & Run-time optimised &  &   &  &  & \checkmark  &  & & \checkmark  & \\ \hline
 
 
 \multicolumn{2}{|c|}{\multirow{5}{*}{ Fog/Edge resource handling} } &  \multirow{2}{*}{Connectivity} & Manual registration & \checkmark & \checkmark  & \checkmark & \checkmark &   &  &  & \checkmark & \\ \cline{4-13}
 
 \multicolumn{2}{|c|}{} &  & Automatic registration &  &   &  &  & \checkmark  & \checkmark & \checkmark & \checkmark & \checkmark  \\ \cline{3-13}
 
 \multicolumn{2}{|c|}{} & \multirow{3}{*}{Others} & Heterogeneity & \checkmark & \checkmark  & \checkmark & \checkmark & \checkmark  & \checkmark & \checkmark & \checkmark  & \checkmark\\ \cline{4-13}
 
 \multicolumn{2}{|c|}{} &  & Auto reconnectivity & \checkmark &   & \checkmark & \checkmark & \checkmark  &  & \checkmark & & \checkmark \\ \cline{4-13}
 
 \multicolumn{2}{|c|}{} &  & Resource discovery &  &   & \checkmark &  & \checkmark  & \checkmark & \checkmark & & \checkmark  \\ \hline
 
 \multirow{28}{*}{\rotatebox{90}{Orchestration functionalities}}  &  \multirow{4}{*}{Service/Job Handling}  & \multirow{2}{*}{Virt support}  & VM &  &  &  & \checkmark &   &  &  & & \\ \cline{4-13}
 
  & & & Containerisation & \checkmark & \checkmark  &\checkmark & \checkmark & \checkmark  & \checkmark & \checkmark & \checkmark & \checkmark   \\ \cline{3-13}
  
   & & \multirow{2}{*}{Mapping} & Static& \checkmark & \checkmark  &  & \checkmark &   &  &  & \checkmark &  \\ \cline{4-13}
 
   & & & Context aware & $\sim$  & \checkmark  &  &  & \checkmark  & \checkmark & \checkmark & & \checkmark \\ \cline{2-13}
 
 & \multirow{11}{*}{Run-time reconfiguration} & \multirow{2}{*}{Definition type} & Statically pre-defined&  & \checkmark  & \checkmark & \checkmark &   & \checkmark & \checkmark & & \\ \cline{4-13}
 & & & User-defined dynamic& \checkmark &   &  & \checkmark & \checkmark  &  &  & \checkmark & \checkmark \\ \cline{3-13}
 
 & & \multirow{3}{*}{Operating type} & Reactive& \checkmark & \checkmark  & \checkmark & \checkmark &   & \checkmark & \checkmark & \checkmark & \checkmark \\ \cline{4-13}
 & & & Proactive&  &   &  &  & \checkmark  &  &  & & \\ \cline{4-13}
 & & & Hybrid&  &   &  &  &   &  &  & & \\ \cline{3-13}
 
 & & \multirow{3}{*}{Scaling} & Horizontal& &   &  & \checkmark & \checkmark  &  &  &  &  \\ \cline{4-13}
 & & & Vertical&  &   &  &  & \checkmark  &  &  &  & \\ \cline{4-13}
 & & & Hybrid &  &   &  &  &   &  & \checkmark & \checkmark &  \\ \cline{3-13}
 
 & & \multirow{3}{*}{Offloading}& Cloud-to-Edge&  & \checkmark  &  &  &   &  &  & \checkmark & \checkmark  \\ \cline{4-13}
 & & & Edge-to-Cloud&  & \checkmark  &  &  &   &  &  & \checkmark & \checkmark  \\ \cline{4-13}
 & & & Edge-to-Edge& \checkmark  &   &  &  & \checkmark  &  & \checkmark & & \checkmark \\ \cline{2-13}
 
 & \multirow{6}{*}{Monitoring} & \multirow{3}{*}{Support level} & Cloud&  &   &  &  &   &  & \checkmark & & \\ \cline{4-13}
 & & & Edge&  &   &  &  &   &  & \checkmark &  & \\ \cline{4-13}
 & & & Cloud-to-Edge & \checkmark & \checkmark  & \checkmark & \checkmark & \checkmark  & \checkmark &  & \checkmark & \checkmark \\ \cline{3-13}
 
 & & \multirow{3}{*}{Metrics support} & System& \checkmark & \checkmark & \checkmark & \checkmark & \checkmark  & \checkmark & \checkmark & \checkmark & \checkmark  \\ \cline{4-13}
 
 & & & Application& \checkmark & \checkmark  & \checkmark & \checkmark & \checkmark  &  & \checkmark & \checkmark & \checkmark \\ \cline{4-13}
 
 & & & Custom& \checkmark & \checkmark  &  & \checkmark & \checkmark  &  &  & \checkmark &  \\ \cline{2-13}
 
 & \multicolumn{2}{c|}{\multirow{4}{*}{Security handling} } & Configurable app level&  &   &  & \checkmark &   &  &  & & \\ \cline{4-13}
 
 & \multicolumn{2}{c|}{}  & Sys wide inter-comp&  &   & \checkmark & \checkmark & \checkmark  &  &  & \checkmark & \\ \cline{4-13}
 
 & \multicolumn{2}{c|}{}  & Edge authentication&  &   &  &  &   & \checkmark & \checkmark & \checkmark &   \\ \cline{4-13}
 
 & \multicolumn{2}{c|}{} & Access control& \checkmark &   &  &  &  \checkmark &  & \checkmark & \checkmark & \\ \cline{2-13}
 
 & \multicolumn{2}{c|}{\multirow{2}{*}{Others}} & Fault diagnosis&  &   &  & \checkmark & \checkmark  &  &  & & \\ \cline{4-13}
 
 & \multicolumn{2}{c|}{} & SLA Handling&  & & $\sim$ &  & $\sim$  & $\sim$ & $\sim$ & $\sim$  & \\ \hline
 
 \multicolumn{2}{|c|}{\multirow{10}{*}{Design} } & \multirow{3}{*}{Architecture} & Centralised& \checkmark & \checkmark  &  & \checkmark & \checkmark  & \checkmark &  & \checkmark & \\ \cline{4-13}
 
 \multicolumn{2}{|c|}{} &  & Decentralised &  &   & \checkmark &  &   &  & \checkmark & & \checkmark \\ \cline{4-13}
 
 \multicolumn{2}{|c|}{} &  & Hybrid &  & &  &  &   &  &  & & \\ \cline{3-13}
    
 \multicolumn{2}{|c|}{} & \multirow{2}{*}{App description} & Solution independent & \checkmark & \checkmark  &  & \checkmark & \checkmark  & \checkmark &  & \checkmark  & \\ \cline{4-13}
 
 \multicolumn{2}{|c|}{} &  & Solution specific &  &   & \checkmark &  &   &  & \checkmark & & \\ \cline{3-13}

 \multicolumn{2}{|c|}{} & \multirow{2}{*}{Extensibility} & Resources & \checkmark &   &  & \checkmark &\checkmark  &  &  & \checkmark & \\ \cline{4-13}
 
 \multicolumn{2}{|c|}{} &  & Functionalities &  &   &  & \checkmark & \checkmark  &  &  & & \\ \cline{3-13}
 
 \multicolumn{2}{|c|}{} & \multirow{3}{*}{User interface} & GUI (Web/Desktop) & \checkmark &   &  &  & \checkmark  & \checkmark & \checkmark & \checkmark & \\ \cline{4-13}
 
 \multicolumn{2}{|c|}{} &  & CLI & &   &  & &   &  & \checkmark & & \\ \cline{4-13}
 
 \multicolumn{2}{|c|}{} &  & API (Service/Library) & & \checkmark  & \checkmark & \checkmark &   &  & \checkmark & \checkmark & \checkmark \\ \hline
 

 \multicolumn{3}{|c}{} & Supported App types & G & DI  & G & G & DI  & SC & G & G & G\\ \hline
\end{tabular}
  {\raggedright [Supported = \checkmark, partially supported = $\sim$]  Supported App types $\Rightarrow$ Generic \textbf{(G)}, Data Intensive \textbf{(DI)}, Smart construction \textbf{(SC)}  \par}
\label{table:research_oriented}
\end{table*}

\textbf{Slack4things}~\cite{slack4things} is an open-source initiative developed by the Mobile and Distributed Systems Lab (MDSLab) at the University of Messina, Italy. This project aimed to provide an OpenStack-based IoT framework for managing IoT devices seamlessly, i.e., without considering their physical location, network configuration, and underlying technology. The tools from this project are further extended by Merlino et al.~\cite{merlino2019enabling} to build a distributed orchestration solution based on a three-layer architecture that covers cloud-fog-edge, and 
supports both horizontal and vertical task offloading. With the former, tasks can be migrated within the same layer, e.g., from one edge device to another; with vertical offloading, tasks can move across different layers, e.g., from edge to fog, or from fog to cloud. Unlike other systems, this solution is based on 
independent managers deployed at each layer of the architecture; hence, 
applications can be directly deployed, partly or as a whole, to any layer through the provided managers.

Beyond the projects introduced above, there are some relevant EU research initiatives that have just recently started; 
however, at the time of review, we were not able to find any reported results from these initiatives therefore, we only briefly review them below for the purpose of completeness. The \textbf{European Cloud, Edge and IoT (CEI) Continuum} \cite{euCloudEdgeIot} is an umbrella initiative that provides the strategic guidance and next stage of tech development to achieve the goals of an active and dynamic European CEI ecosystem, with an emphasis on promoting the establishment of a global and open ecosystem for the Cloud-Edge-IoT technologies. 
%
The initiative coordinates across clusters of Research and Innovation Actions to support industries and researchers in creating impact, promoting the link between open source and open standards, and engaging relevant industrial alliances in actions directed toward open approaches.
Among such clusters, the Meta-Operating Systems for the Next Generation IoT and Edge Computing (MetaOS) \cite{metaOs} are relevant for this review paper, and include projects such as AerOS, FluiDOS, ICOS, NebulOus, NEMO and NEPHELE. 
Likewise, the cluster AI-enabled computing continuum from Cloud to Edge (CognitiveCloud) \cite{cognitiveCloud} is also related to our work, and includes projects such as AC3, ACES, CloudSkin, CODECO, COGNIFOG, DECICE, EDGELESS, MLSysOps and SovereignEdge.Cognit. More details of these projects are available on the CEI website.
\paragraph{\textbf{Industry initiatives}}
This section presents an overview of some of the existing industry platforms that support the combined orchestration of cloud and edge resources. The key factors that led to the inclusion of these solutions are both the availability of their implementation and the presence of technical documentation and/or associated white papers. It is important to note that, even though such solutions often rely on underlying open source components, such as Docker and Kubernetes, they are in fact vendor specific, with their scope mostly focused on the orchestration of (5G) network services. Moreover, these being industry-oriented solutions, we found that they often lacked documentation providing in-depth descriptions of the related technical details. Instead, the available documentation focused more on the presentation of features for targeting customers. Therefore, the evaluation of the characteristics of these solutions against the taxonomy was not obvious due to the lack of information. Nonetheless, we include these solutions in the paper for the purposes of completeness, even though a detailed comparative summary table will not be presented in this section.

\textbf{HPE GreenLake}~\cite{hpe-greenlake} cloud-to-edge is an infrastructure-as-a-service solution that brings the public cloud model to multiple IT environments, such as private clouds, multi-cloud and on-premises, in order to deliver an agile cloud everywhere modality to the users. HPE GreenLake allows for the integration, management and monitoring of all the above resources through a centralised interface. Users can access different types of deployable resources and services, e.g., bare-metal, compute, storage, containers and data protection services, as well as HPC, AI/ML and virtual desktop infrastructures.

\textbf{Intel Smart Edge Open}~\cite{intel} is an edge computing software toolkit for building platforms optimized for the edge. This is done by providing a toolkit of functionality selected from across the cloud native landscape, which has been extended and optimised to be used at the edge. This solution is able to work with heterogeneous hardware resources from the on-premise edge to regional data centres. These are managed by using a set of ``experience kits'', provided by Intel and built on top of Kubernetes, that combine 5G capabilities and cloud-native components to simplify the deployment of complex network architectures, significantly reducing development time and cost. For instance, the Developer Experience Kit provides the base capabilities to run containerised edge services, including networking, security, and telemetry. An experience kit consists of building blocks that can be chosen according to the customer's needs. Specifically, resource management provides identification, configuration, allocation, and continuous monitoring of the hardware and software resources on the edge cluster; the Telemetry and Monitoring combine application telemetry, hardware telemetry, and events to create a heat-map across the edge cluster and enable the orchestrator to make scheduling decisions.

\textbf{AMCOP}~\cite{amcop}---Aarna Networks Multi Cluster Orchestration Platform---is an open-source platform for orchestration, life-cycle management, and closed-loop automation of cloud-native network services and edge computing applications. AMCOP aims to solve the problem of managing the growing number of edge applications and edge sites by offering intent-based orchestration of network services and composite edge computing applications, which comprise cloud-native network functions and cloud-native applications; it also supports service assurance for edge and 5G services through real-time, policy-driven closed-loop automation. AMCOP works by interfacing (northbound) with the collection of systems/applications that a network service provider already uses to operate its business (OSS/BSS), and by orchestrating infrastructure and network services/applications across multiple heterogeneous Kubernetes clusters (southbound).

\textbf{Ormuco}~\cite{ormuco} is a solution that aims to lead the deployment and usage of edge computing as an effective approach to deliver data processing. The platform was developed to respond to the needs of modern businesses that require the setup of an infrastructure-as-a-service via a decentralised approach in order to increase their revenue, reduce both the operations and maintenance costs, and automate the deployment of systems and applications on demand. The platform's Cerebro virtual sysadmin is able to collect logs from heterogeneous computing nodes and applications; these are used to learn the expected behaviour of the deployed software and to notify application owners of any detected potential anomalies.

\textbf{Azion}~\cite{azion} is an end-to-end encrypted edge orchestration service with cloud management and zero-touch provisioning, created for large-scale edge networks. Users can manage and control resources across the edge in real-time and orchestrate services more easily, according to specific service requirements. The orchestration relies on an agent, installed on the edge nodes, that provides encrypted remote node management to the Azion Control panel, within the Real-Time Manager, deployed in the cloud. The Edge Node module enables devices to be created and managed and implements the integration with the orchestrator. The Edge Services module enables the customers to create their own services and allows them to be managed and orchestrated by the Real-Time Manager.

\textbf{ONAP}~\cite{onap} platform provides a unified operating framework for vendor-agnostic, policy-driven service design and implementation, as well as analytics and lifecycle management for large-scale workloads and services. Network operators can use ONAP to orchestrate both physical and virtual network functions; hence, they can capitalise on their existing network infrastructure while being part of a vibrant VNF ecosystem that includes providers around the globe. The ONAP Operations Manager (OOM) module, based on Kubernetes is responsible for orchestrating the end-to-end lifecycle management and monitoring of ONAP components, as well as enforcing scalability and resiliency mechanisms.

\textbf{ZEDEDA}~\cite{zezeda} is a cloud-based orchestration solution for the secure control of distributed edge computing deployments, which provides users with full-stack remote management of edge computing hardware and applications deployed both on clouds or on-premises systems. ZEDEDA leverages EVE-OS~\cite{eve}, a secure, open universal operating system, developed with vendor-neutral and open-source governance as part of the Linux Foundation's LF Edge organization. EVE-OS simplifies the deployment, orchestration and security of cloud-native and legacy applications on distributed edge compute nodes. EVE-OS encrypts data, maintains device and software integrity and supports VMs, containers and clusters (Docker and Kubernetes).

\section{Discussion, Issues, and Future Directions}
\label{sec:discussion}
Section~\ref{sec:review} thoroughly reviewed the existing solutions by classifying them into different groups. Table~\ref{table:conceptOnly} and~\ref{table:research_oriented} further summarised the solutions from the concept-only and research initiatives categories by outlining their characteristics based on the proposed taxonomy. This section discusses and reflects on the results from tables with the aim of highlighting directions for future work in relation to the advancement of cloud-to-edge orchestration. More particularly, Section~\ref{sec:issues} discusses the open issues that require further consideration, whereas Section~\ref{sec:framework}, based on the analysis of the CoTOS landscape and the identification of notable gaps, presents a generic high-level conceptual framework for the development of the next generation CoTOS.

\subsection{Open Issues}\label{sec:issues}
\subsubsection{Standardised support for application description}
A key distinction that we made in this paper is the differentiation between lower-level and higher-level  solutions using the presence (or lack) of a high-level abstraction layer. Such an abstraction aims to increase portability and interoperability by empowering users to specify their applications' requirements using a high-level standardised method to avoid any kind of vendor and/or technology lock-in. It is evident from results by focusing on the \textit{``App description''} attribute that a number of solutions such as ENORM~\cite{wang2017enorm}, Fernandez et al~\cite{fernandez2019enabling}, Alam et al.~\cite{alam2018orchestration}, Santos et al.~\cite{santos2017fog} do not provide such an abstraction layer. A number of other solutions, including Foggy~\cite{foggy}, Gabriele et al.~\cite{castellano2019service}, mF2C~\cite{masip2021managing}, and Rainbow~\cite{rainbow-project} used a YAML or custom DSL based abstraction mechanism. However, all these approaches are specific to respective solutions, hence solution dependent. On the other hand, most of the solutions in the research initiative category including SODALITE@RE~\cite{sodalite}, Capillary~\cite{capillary}, MiCADO-Edge~\cite{ullah2021micado}, PrEstoCloud~\cite{verginadis2021prestocloud, verginadis2017prestocloud}, DECENTER~\cite{kochovski2021building} and Pledger~\cite{pledger-project} follow a solution independent approach, 
where they mainly apply the TOSCA standard format for application description. TOSCA is a well-known, popular and standardised cloud orchestration modelling language that has been extensively used by many cloud orchestration tools. However, the latest TOSCA standard (i.e., Version 1.3~\cite{toscaLatestStandard} at the time of writing) still lacks support for edge-related aspects. Due to the lack of native support, all the aforementioned solutions provide ad-hoc extension and implementation for the edge-related aspects. Such ad-hoc adoption of TOSCA loses the inherent portability aspects of utilising TOSCA as an abstraction layer. Hence, further efforts are required to develop (or extend) existing standardised modelling languages to enable native support for edge-related aspects.

\subsubsection{SLA management} 
The complex nature of the cloud-to-edge continuum, which consists of heterogeneous computational and communication infrastructures belonging to multi-domains, and often relies on ephemeral mobile, low-power computational devices with volatile connectivity, brings in possibly varying run-time conditions that can ultimately influence the delivery of the expected QoS. Therefore, from a system viewpoint, the management of SLAs of IoT applications is a very complex task. It is evident from the results by referring to the \textit{``SLA handling''} attribute that, some of the existing works, e.g., ENORM~\cite{wang2017enorm}, Fernandez et al~\cite{fernandez2019enabling} from the concept-only category, and almost all solutions from the research initiative category have considered the SLA aspect. However, the majority of these solutions have focused only on SLA enforcement, where they perform reconfiguration as a result of the violation of certain conditions. We consider that the scope of SLA management is larger than just SLA enforcement, and orchestration systems are also needed to focus on other related aspects, e.g., a standardised way of describing SLA, reporting of SLA violation, formal assurance of SLA, SLA negotiation considering different administrative domains, and SLA monitoring across different administrative domains. However, amongst the reviewed solutions, very limited efforts are provided with regard to these aspects. Some limited notable examples include the reporting of SLA violations by mF2C~\cite{masip2021managing}, and the formal assurance of SLA in DECENTER~\cite{kochovski2021building}. For the SLA specification, which is an essential task for QoS-aware orchestration, a number of solutions provide their custom method to define a reconfiguration policy, e.g., MiCADO-Edge asks application owners to define their SLA by writing a Python-based scaling policy using a set of exposed variables, while mF2C allows resource based conditions. However, all such methods assume specific knowledge of the underlying systems. On the other hand, commercial solutions, such as AMCOP and ONAP also have their SLA specification mechanisms based on existing ETSI proposals~\cite{etsi-SLA}. However, the scope of these is mainly focused on the network domain only and hence may not be applicable to the whole cloud-to-edge continuum. Further efforts are required to provide a standardised format that facilitates application owners to specify the SLA requirements of their applications. Some limited individual works in this direction include~\cite{alqahtani2018end,antonescu2015service}. These works advocate the use of model-driven SLA specifications that extend TOSCA with SLA-defining constructs. For other aspects, there are also some initial works, e.g., for QoS negotiation~\cite{antonacci2019digital, costantini2021cloud}, and for formal assurance of SLA~\cite{kochovski2020smart,alzubaidi2019blockchain}. From an overall orchestration viewpoint, these aspects are essential and there should be more focus on incorporating these functions.

\subsubsection{Context-aware resource discovery} 
As already pointed out for SLA Management, the multi-domain, heterogeneous nature of the cloud-to-edge continuum poses considerable challenges to the unified management of the available resources. These should ideally be made available as a pool that can grow and shrink as resource elements are dynamically discovered across the different layers of the involved administrative domains. The results (relevant attributes include: \textit{``Resource discovery''} and \textit{``Mapping''}) inform us that most of the solutions lack a resource discovery mechanism and therefore provide a static mapping, where users manually define the association of components to specific resources, rather than a context-aware one. However, the cloud-to-thing continuum model is highly dynamic, where IoT applications have different requirements relating to resources, e.g., locality, type of resources, operating system, architecture, and remaining battery power. Hence, from a uniform resource management viewpoint, an orchestration solution should provide a standardised dynamic way to empower resource owners to securely register their resources using the various associated contextual attributes, and where these resources can be dynamically discoverable across different administrative domains at the time of deployment and reconfiguration decisions. Hence, we foresee the usage of solutions similar to the one described in our previous work \cite{tusa-slices}, where a \textit{Resource Marketplace} was considered as a dynamic approach for sharing resource availability information among various domains. This also allows the execution of mapping algorithms aimed at identifying resources across the cloud-to-edge continuum that can be used to satisfy the performance requirements of the IoT services requested by the users.

\subsubsection{Proactive run-time reconfiguration} It is evident from the results by focusing on the \textit{``Operating type''} attribute that almost all of the reviewed solutions are reactive in nature, rather than proactive. Such solutions execute reconfiguration actions (i.e., either scaling and/or offloading) in response to the changes in the behaviour of the system that can be identified through the fulfilment (or violation) of some conditions. The key issue with the reactive approach is the delay elapsed between the time of the reaction to a change and the actual completion time of the reconfiguration process~\cite{ullah2017towards}. Reactive orchestration is also more prone to oscillation. You may react too fast and then react again, causing too many reconfigurations. The proactive approach on the other hand anticipates future behaviour of the system and performs the necessary reconfiguration in advance. Within the cloud-only domain, there is already a lot  of attention provided on the use of proactive approaches to perform auto-scaling (e.g., please check these review papers for further details~\cite{lorido2014review, ullah2018control}). However, as evident from the results, there is very limited attention on the use of proactive approaches by orchestration systems in the cloud-to-edge continuum. Therefore, a new breed of machine learning-based contextual models needs to be developed to make reconfiguration decisions by anticipating future system behaviour considering a large range of relevant contextual information, such as System resources (e.g., utilisation, capabilities, types, availability), Networking aspects (e.g., congestion level, available bandwidth, communication overhead), Energy requirements (e.g., battery utilisation, battery life), Environmental context (e.g., locality, time of the day), Application service context (low latency, QoS specification), and Social behaviour aspects, (e.g., roaming habits of users). 

\subsubsection{Decentralised architecture} Most of the reviewed solutions, especially the ones with implementations, follow a centralised approach. This approach offers a number of key advantages, such as significantly reducing the complexity of the implementation and offering consistent decision-making, thanks to the information relevant to the decision process being located in a single place. However, the centralised approach also comes with significant drawbacks~\cite{hong2019resource}. For example, it suffers from the lack of scalability, it offers a single point of failure and a centralised target for cyber-attacks. Moreover, a single central component can easily become a bottleneck from an efficiency point of view. Such a model fits more naturally in the single-cloud domain. However, considering the cross-cloud and distributed nature of the cloud-to-edge ecosystem, a centralised approach also raises additional concerns, e.g., the transfer of recurrent monitoring data from distributed resources to a central point, sporadic connectivity, privacy and locality constraints. A decentralised approach can address the aforementioned limitations. In such an approach, multiple orchestrators (decision makers) work independently, and each orchestrator manages its dedicated applications or its own domain of the infrastructure and also collaborates with each other to reach QoS standards or fulfil SLAs or policy requirements \cite{5GEx, tusa-mano}.

\subsubsection{Security handling} The core focus of this paper was not on the security aspects of orchestration systems. However, we still evaluated existing solutions in four basic but essential aspects related to security. It is evident from the results by focusing on the sub-attributes of \textit{``Security handling''} that only a few solutions have very limited and partial attention to some of the four aspects. Hence, more attention is required in this regard. One of the key challenges for orchestration systems related to security is the resource-constrained computational devices at the edge, which in some cases are unable to support the traditional security methods. Therefore, new methods need to be designed considering the distributed, multi-administrative domain and resource-constrained nature of the cloud-to-edge environment. Further specific details in relation to security challenges and some solutions can be found in~\cite{ren2019survey,fakude2019fog,vsatkauskas2020orchestration}.
\subsection{A conceptual framework of orchestration in the Cloud-to-Things compute continuum} \label{sec:framework}
Based on the analysis of the CoTOS landscape, and the identification of notable gaps described in the previous section, we present a generic high-level conceptual framework for the development of the next generation CoTOS. This section describes the high-level abstract details of the framework to help researchers and developers in identifying potential components and building blocks, based on the functionalities and notable missing features of the current solutions. 

The ideas that underpin our framework proposal are also aligned with the objectives targeted 
by some of the newly started research initiatives in the context of orchestration in the cloud-to-edge continuum, such as those mentioned earlier in the paper \cite{euCloudEdgeIot}. Similar to \cite{metaOs}, we also foresee the development of a unified framework for smart IoT applications, acting as a Meta-Operating System that enables seamless cloud and edge computing orchestration by bringing computation, data and intelligence closer to where the data is produced. As in \cite{cognitiveCloud}, AI techniques will also be used to build a cognitive framework that will automatically adapt to the growing complexity and data deluge by integrating seamlessly and securely diverse computing and data environments, spanning from core cloud to edge. 

The implementation aspects of the suggested framework's components and overall execution model are left out for developers to choose based on their key requirements. Figure~\ref{fig:framework} illustrates the high-level architecture of our envisioned system, where the overall structure has been categorised into three layers, namely the \textit{Application Description layer}, the \textit{Orchestration layer}, and the \textit{Infrastructure layer}. The following sections explain each of these layers.

\subsubsection{Application Description layer} We expect enabling application owners to produce vendor-agnostic and interoperable deployment models for their applications that will be deployed, managed and executed within the Cloud-to-Things continuum. We aim to have a single uniform standardised deployment descriptor (a deployment model for the application) that incorporates all the relevant details of the target application, including the topological structure of the involved microservices, the specification of the cloud-to-edge resource requirements and SLA specification, reconfiguration, and application-level security policies. Inspired from~\cite{tsagkaropoulos2021extending, tusa-slices}, we consider a multi-level deployment model. A \textit{high-level}, where an application owner provides an abstract description of the required resources and some optimisation criteria, rather than specific details in relation to resource providers, and/or instances. Based on user-provided optimisation criteria, the \textit{Cloud/Edge Offerings Manager} component is responsible for producing an optimised deployment model consisting of specific details of the resources that will be used. The optimised deployment model is then passed on to the \textit{Deployment Model Manager} component. 
Alternatively, application owners could also follow a static approach, whereby they can directly produce a deployment model, containing all the required resource-specific details, which can be directly passed on to the \textit{Deployment Model Manager} component. 

\subsubsection{Orchestration layer} 
\begin{figure*}[!ht]
\includegraphics[scale=0.5]{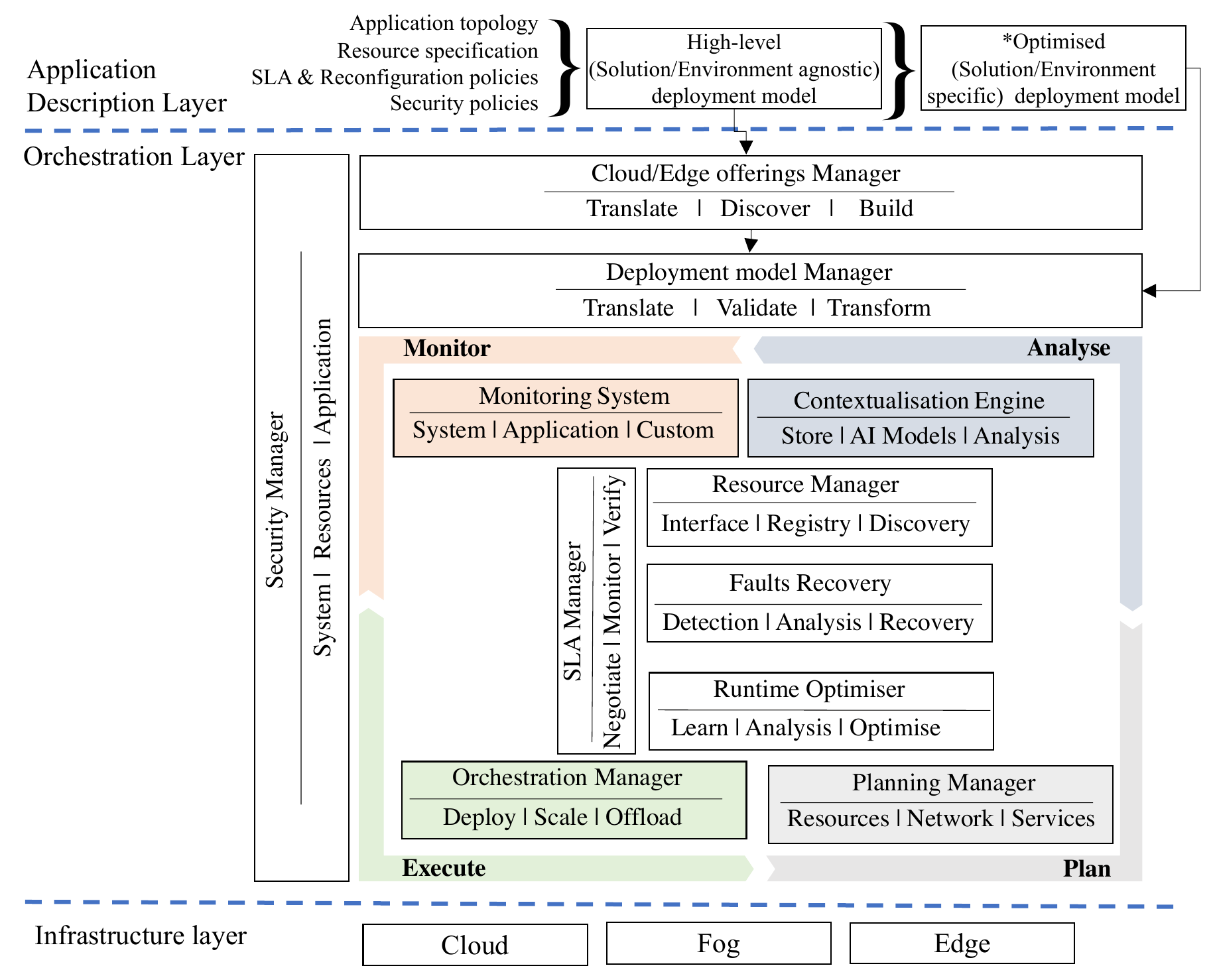}
\caption{Conceptual framework of orchestration in the Cloud-to-Things computing continuum}
\label{fig:framework}
\end{figure*}

At this layer, the \textit{Deployment Model Manager} receives either the \textit{high-level} or \textit{intermediate-level} deployment model mentioned above, and it is responsible for the translation, validation and transformation of the model into the corresponding low-level details related to the platform (and resources), such as specific orchestration manifests, policies enforcement. For the key operations, i.e., the deployment, reconfiguration and run-time operational management of the target application, we envision the MAPE-K~\cite{computing2006architectural} loop architectural concept of self-adaptive systems. As Figure~\ref{fig:framework} depicts, there are specific components that are responsible for each stage of the MAPE-K loop. More particularly, the role of the \textit{Monitoring System} is to collect the system and application-related metrics, as well as the related contextual information, across the entire Cloud-to-Things ecosystem. The collected information will be received by the \textit{Contextualisation Engine} 
and utilised at different points in time; 
It will be fed into a wide range of relevant Artificial Intelligence models to make system-level proactive decisions related to the applications being executed. These decisions will be further planned and scheduled in terms of specific actions, considering the available resources by the \textit{Planning Manager}. Lastly, the \textit{Orchestration Manager} will execute the planned actions.

The key MAPE-K components of the envisioned system are supported by the \textit{Resource Manager}---responsible for the overall management of both cloud and non-cloud resources. For the non-cloud resources, it will also manage resource registration and dynamic discovery at run-time; \textit{Run-time Optimiser}---responsible to optimise the overall deployment setup in terms of resource usage and application performance; \textit{SLA Manager}---responsible to manage the SLA related aspects, including providing an interface for negotiation, monitoring and verification at run-time; and the \textit{Faults Recovery}---responsible for detecting run-time errors in relation to application services and resources, and taking steps for automatic recovery. Lastly, the \textit{Security Manager} consists of various enablers that are responsible for dealing with the overall security of the system across the multi-level cloud-to-edge layers, such that the entire system can securely operate 
on heterogeneous resources geographically dispersed across multiple domains. It will also be responsible for enforcing the application-level security policies defined by the application owners at the \textit{Deployment layer}. Moreover, the \textit{Security Manager} should also provide methods for authentication of resources (and IoT devices), such that unauthorised use of resources can be eliminated at the resource level and not only at the system component level.

\subsubsection{Infrastructure layer} 
Various types of geographically dispersed resources belonging to the Cloud-to-Things continuum, and potentially spanning multiple administrative domains, are the constituting elements of this layer. These resources will be used by the above-mentioned components of the \textit{Orchestration layer} in order to simultaneously deploy, reconfigure and manage IoT applications.

\section{Conclusion}\label{sec:conclusion}
The increasing adoption of the Cloud-to-Things computing model emphasises the importance of intelligent and robust orchestration solutions to address the quintessential needs of modern IoT applications, which require simultaneous access and management of geographically distributed arrays of sensors, heterogeneous remote, local and multi-cloud computational resources, as well as dynamic handling of the application execution. In this paper, we thoroughly reviewed a diverse range of existing orchestration solutions; we then proposed a novel taxonomy that consists of a wide set of characteristics that we deemed essential for the automated deployment and run-time management of IoT applications within the Cloud-to-Things continuum.

Based on the obtained results from this review, we identified six key areas, where current solutions are lacking focus. These areas include standardisation support for application description, SLA management, context-aware resource discovery, proactive run-time reconfiguration, decentralised architectures, and security management. These areas highlight directions for future work. Moreover, based on these identified areas, we also presented a proposal for a conceptual framework that can provide a foundation for the implementation of future orchestration solutions.


\begin{backmatter}

\section*{Funding}
This work was supported by the following projects funded by the European Commission's H2020 Programme: DIGITbrain (project number  952071), PITHIA-NRF (project number 101007599) and Harpocrates (project number 101069535).

\section*{Competing interests}
The authors declare that they have no competing interests.

\section*{Authors' contributions}
All authors equally contributed to the design of taxonomy and classification and the general conceptualisation and concept of the paper. \\
\textbf{Amjad:} Initial conceptualisation, Design of the conceptual framework, Contributed to the write-up of Sections including Introduction, Related Work, Taxonomy, Concept-only, Research initiatives, “Discussion, issues, and future directions" and "A proposal of conceptual framework". \\
\textbf{Tamas:} Design of the conceptual framework, and review of the entire manuscript. \\
\textbf{József:} Contributed to the design of taxonomy, the write-up of the Section "Industry initiatives", and review of the entire manuscript. \\
\textbf{Francesco:} Contributed to the write-up of sections including  "Industry initiatives", “Discussion, issues, and future directions", "A proposal for a conceptual framework", and review of the entire manuscript. \\
\textbf{James:} Contributed to the write-up of Section "Lower-level solutions". \\
\textbf{Huseyin:} Contributed to the write-up of Section "Lower-level solutions", and review of the entire manuscript. \\
\textbf{Resmi:} Contributed to the write-up of Section "Lower-level solutions". \\
\textbf{Hamed:} Contributed to the write-up of Section "Research initiatives".
\section*{Declarations}

\section*{Ethical Approval}
Not applicable

\section*{Availability of data and materials}
Not applicable

\bibliographystyle{bmc-mathphys} 
\bibliography{main}      







\end{backmatter}
\end{document}